\documentclass[a4paper,11pt]{article}

\usepackage{graphics,graphicx}
\usepackage{natbib}
\usepackage{txfonts}

\usepackage{color}
\definecolor{kleur}{rgb}{1.0,0.08,0.45}
\definecolor{rood}{rgb}{1.0,0.08,0.08}
\definecolor{blauw}{rgb}{0.08,0.08,1.0}

\newcommand{\blue}[1]{\textcolor{blauw}{#1}}

\newcommand{\stress}[1]{{\blue{\textbf{\textit{#1}}}}}

\usepackage{xspace}
\newcommand{\hirex}{\textsl{HiReX}\xspace}
\newcommand{\rgs}{\textsl{RGS}\xspace}
\newcommand{\xmm}{\textsl{XMM-Newton}\xspace}
\newcommand{\chandra}{\textsl{Chandra}\xspace}
\newcommand{\letgs}{\textsl{LETGS}\xspace}
\newcommand{\hetgs}{\textsl{HETGS}\xspace}
\newcommand{\suzaku}{\textsl{Suzaku}\xspace}
\newcommand{\xrism}{\textsl{XRISM}\xspace}
\newcommand{\resolve}{\textsl{Resolve}\xspace}
\newcommand{\hitomi}{\textsl{Hitomi}\xspace}
\newcommand{\einstein}{\textsl{Einstein}\xspace}
\newcommand{\exosat}{\textsl{EXOSAT}\xspace}
\newcommand{\athena}{\textsl{Athena}\xspace}
\newcommand{\xifu}{\textsl{X-IFU}\xspace}
\newcommand{\asca}{\textsl{ASCA}\xspace}
\newcommand{\sax}{\textsl{Beppo-SAX}\xspace}

\newcommand{\ion}[2]{#1\,{\sc{#2}}}

\usepackage{abstract}

\usepackage[font=footnotesize]{caption}

\makeatletter
\renewcommand{\section}{\@startsection%
{section}{1}{0mm}{-\baselineskip}%
{0.5\baselineskip}{\normalfont\Large\bfseries}}%
\makeatother

\begin{document}
\pagestyle{plain}
\pagenumbering{arabic}


\begin{titlepage}
 \begin{center}
 \vspace*{1cm}
 

{\huge \textbf{The Voyage of Metals in the Universe from Cosmological to
Planetary Scales}}

\vspace{0.5cm} 
{\Large the need for a Very High-Resolution, \\ High Throughput Soft
X-ray Spectrometer}

\vspace{0.5cm}
\begin{figure*}[!h]
\centering
\includegraphics[width=13.7truecm]{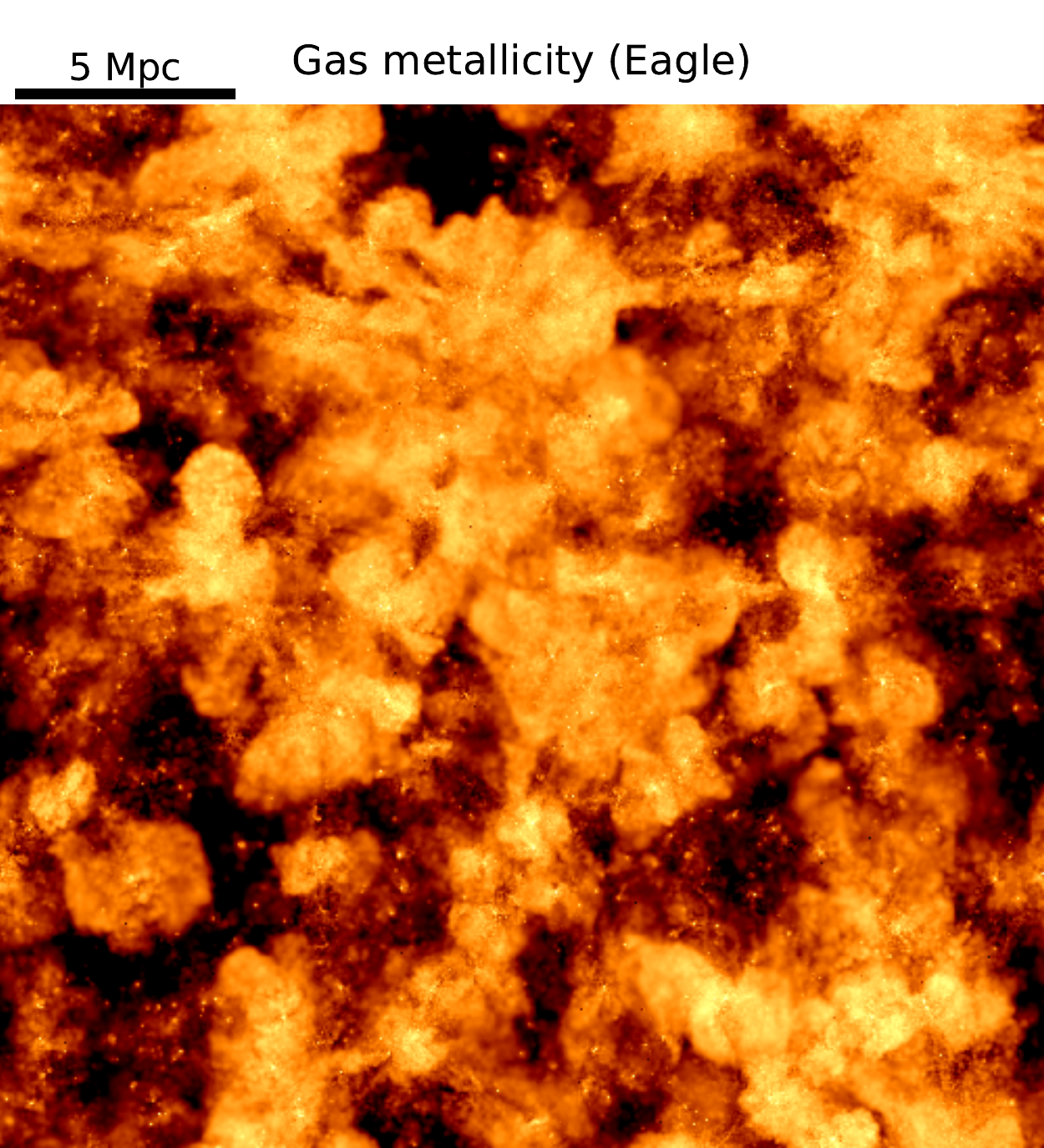}
\end{figure*}

 \vspace{0.4cm}
 
 \textbf{F. Nicastro$^1$ \& J. Kaastra$^2$}
 
 \vspace{0.3cm}
 \footnotesize{
 $^1$ Istituto Nazionale di Astrofisica - Osservatorio Astronomico di Roma (INAF - OAR), Via di Frascati 33, 00078 Monte Porzio Catone (RM), Italy. 
\newline
 $^2$ Netherlands Institute for Space Reasearch (SRON), Sorbonnelaan 2, 3584 CA Utrecht, The Netherlands. 
}

 \end{center}
\end{titlepage}




\begin{abstract}  

\normalsize 
Metals form an essential part of the Universe at all scales. Without metals we
would not exist, and the Cosmos would look completely different. Metals are
primarily born through nuclear processes in stars. They leave their cradles
through winds or explosions, and then start their journey through space. This
can lead them in and out of astronomical objects on all scales, ranging from
comets, planets, stars, entire galaxies, groups and clusters of galaxies to the
largest structures of the Universe. Their wanderings are fundamental in
determining how these objects, and the entire universe, evolve. In addition,
their bare presence can be used to trace what these structures look like.
\smallskip

The scope of this paper is to highlight the most important open astrophysicals
problems that will be central in the next decades and for which a deep
understanding of \stress{the Universe-wandering metals, their physical and
kinematical states and their chemical composition} represents the only viable
solution.  The majority of these studies can only be efficiently performed
through High Resolution Spectroscopy in the soft X-ray band. 

\end{abstract}

\section{Introduction}

Metals\footnote{All atoms heavier than hydrogen and helium in the common
astronomical use of the word} form an essential part of the Universe at all
scales. Without metals we would not exist, and the Cosmos would look completely
different. Metals are primarily born through nuclear processes in stars. They
leave their cradles through winds or explosions, and then start their journey
through space. This can lead them in and out of astronomical objects on all
scales, ranging from comets, planets, stars, entire galaxies, groups and
clusters of galaxies to the largest structures of the Universe. Their wanderings
are often fundamental in determining how these objects evolve. Additionally,
their bare presence can be used to trace what these structures look like.

Because the majority of the ordinary matter in the Universe (i.e. baryons) is in
the form of hot or warm plasma, it emits and absorbs X-ray line radiation that
can only be studied using high-resolution X-ray spectroscopy. But even colder
gas or dust (i.e. the major tracers for star formation) can be studied through
absorption of background X-rays.

After the first pioneering observations with the grating spectrometers of the
\einstein and \exosat satellites, the field of high-resolution X-ray
spectroscopy became mature through the grating spectrometers on board of
\chandra and \xmm (both launched in 1999). In 2016 the first microcalorimeter
spectrum was obtained through the \hitomi satellite. This offered the
opportunity to study spatially extended sources, in particular with high
resolution at high energies. The technique holds great promise for the \hitomi
successors \xrism and ESA's big flagship mission \athena. 

Still, in the soft X-ray band these new missions have lower energy resolution
($R<200$ for oxygen) than the current gratings on \chandra and \xmm, which have
an effective area in this band of 80~cm$^2$ or less. Despite this, the soft
X-ray band below 1.5~keV is actually the richest one in terms of diagnostic
power because it contains spectral lines from all metals heavier than boron in
all ionisation stages.

There is a large number of important astrophysical questions that need high
resolving power in the soft X-ray band, in order to be properly addressed: 

\begin{itemize}

\item{} \stress{Where and in what physical state are the Universe's missing
baryons, and how does their wandering in-and-out of structures affect the
evolution of the Universe and of its different components?}

\item{} \stress{How do winds from Active Galactic Nuclei (AGNs), Supernovae
(SNs) and X-ray binaries (XRBs) redistribute metals in their surroundings?}

\item{} \stress{Where are the metals and in which atomic -- solid state are they
locked?} 

\item{} \stress{How do stellar winds affect the chemical composition of exoplanets and
the general conditions for the existence of life as we know it?}

\end{itemize} 
 
While \athena will open up the road to the solution of these important
problems, by detecting, both in emission and absorption, the densest 20\% of
the extremely tenuous and diffuse gas that permeates the space between and
around galaxies, none of these can be completely addressed without the soft 
X-ray sensitivity (square root of effective area and resolving power) needed to
detect the remaining 80\% of this medium and the resolving power needed to
disentangle all its different physical components and study their dynamics. 
This is key to understanding structure formation and their evolution.

In this paper we present these questions in some detail and conclude
(Sect.~\ref{sect:concept}) by defining the mission-concept needed to complete
\athena's baryon census and address all these questions: a medium-size
(according to ESA standard) soft X-ray dispersive spectrometer (\hirex) with a
resolving power of 5\,000--10\,000, and an effective area of
1\,500--2\,000~cm$^2$ over the 0.2--1.5~keV band. Such a mission in the
2035--2050 timeframe, would not only greatly benefit from its large predecessor
\athena (which will also define a number of important signposts to be followed
up in greater detail by \hirex) but would synergetically operate with all the
already foreseen ground-based as well as space missions over the entire
multi-messenger spectrum.

\section{Galaxy-IGM coevolution: physics, kinematics and chemistry of large
scale inflows and outflows}

During the Universe's childhood (from age $\sim 0.2-2$ billion years), most of its 
baryonic matter, still in an almost primordial composition, permeated the Intergalactic 
medium (IGM), filling the space between gently forming galaxies, nurturing them and in turn receiving
heating photons from newly born stars and the first quasars. These baryons
imprint a forest of \ion{H}{i} Lyman-$\alpha$ absorption lines in the optical
spectra of high-$z$ quasars, and this is how we know of their presence, amount,
location and physical state \citep[e.g.,][]{Rauch98,Weinberg97}. At the age of
only $\sim$ 2 billion years, however, puberty impetuously bursted in and the
Universe's growth became frantic: structures began growing quickly in size, by
devouring material from the surrounding space at higher and higher rates,
phagocytising nearby companions and grouping with close friends. 

According to hydro-dynamical simulations for the formation of structures in the
Universe \citep[e.g.,][]{CenOstriker06,Schaye15}, this activity was accompanied
by a metamorphosis  of the tenuous photo-ionised material filling the space
between galaxies and feeding their growth: baryons in the IGM were more and more
violently pulled towards the growing gravitational  potential wells of
virialised structures and shrunk into a web of sheets and filaments getting
shock-heated to temperatures of $T\simeq 10^5-10^7$ K and so becoming virtually
invisible in \ion{H}{i} absorption. At the same time, freshly metal-enriched
baryons started roaming out of galaxy's disks, pushed out by powerful supernovae
and Active Galactic Nuclei (AGN) winds, wandering into and metal-polluting the circum-galactic 
medium (CGM) and the surrounding IGM. This cycle of baryons and metals in and out of
galaxies proceeded essentially unchanged till our day, and most of the
Universe's ordinary matter today should therefore be in a highly ionised state, heavily
metal-enriched and concentrated in the filaments and nodes of the cosmic web. 

All this however, has not been observationally verified yet (from which, the so called 
'missing-baryon' problem), the main reason being the lack of proper instrumentation in 
the X-rays.  Extreme-UV (to which we are blind because we are located in the disk of the Galaxy) and X-ray photons
are those that most efficiently interact with highly ionised metals, i.e. with
the majority of the baryons of the Universe, so that the physics, kinematics
and  chemistry of these baryons can only be studied through high-resolution
X-ray spectrometers, especially in the 0.1--1.5 keV band, where most of the
astrophysically abundant elements have their  photo-electronic transitions and 
imprint remarkable emission and absorption lines. 

\subsection{The Intergalactic Medium at $z<2$\label{sect:igmz2}} 

The local IGM should be populated with hot baryons with temperatures of about 1
million degrees. At these temperatures, hybridly ionised gas (i.e.\ shock-heated
gas undergoing additional photo-ionisation by the metagalactic radiation
field) has only one neutral atom of hydrogen out of ten million and residual
opacity due to astrophysically abundant metals. In particular the He-like ion of
oxygen is expected to imprint a ``savannah'' of He-$\alpha$ absorption lines,
each with rest-frame equivalent width EW$\lesssim$10 m\AA, in high resolution
soft X-ray spectra of background quasars, analogous to the HI Lyman-$\alpha$
forest seen in the optical spectra of quasars at $z > 2$
\citep[e.g.,][]{CenFang06}. The strongest of these lines (EW $\sim 10-20$ m\AA) are
imprinted by the innermost parts of the CGM of large and rare intervening halos,
while the weakest and more common lines (EW $\sim 1$ m\AA) are produced by
tenuous intervening filaments of the intergalactic web
\citep[e.g.,][]{Fang02,Wijers19}. 

The search for the \ion{O}{vii}-savannah from the missing hot baryons in the
intergalactic medium (the so called Warm-Hot Intergalactic Medium: WHIM) started
about 20 years ago, when the first relatively high resolution X-ray
spectrometers became available with the NASA and ESA observatories \chandra and
\xmm. However, after a number of failed attempts and controversial tentative
results \citep[e.g.,][and references therein]{Nicastro17,Bonamente16} it became
clear that the resolving power and effective areas of these instruments were
just not up to the task and that smart observational strategies (very long
exposures of bright, relatively high-$z$ targets, coupled with the
identification of WHIM signposts) needed to be put in place. 

To date, we can only rely on a couple of possible detections, at locations with
dramatically different surrounding galaxy environments \citep{Nicastro18}. With
current instrumentation, and over the next 10 years, carefully selected
observing programs can be used to strengthen our confidence in these detections
and perhaps add a few more WHIM candidates to the census, but clearly the first 
breakthrough in this field must await missions like \athena, for which one of
the science goals is indeed the detection of the missing baryons in a WHIM.
However, given its limited spectral resolution in the soft X-ray band
(R$\simeq 200$ at 0.5 keV), \athena will only be able to detect 100--200 of the
strongest (EW$\gtrsim$5\,m\AA) \ion{O}{vii} WHIM filaments, which should
represent only about 20\% of the WHIM baryon mass. Moreover, with its $\sim
1500$~km\,s$^{-1}$ resolution at 0.5 keV, \athena will not be able to resolve
the several expected phases of the IGM web, the kinematics (and thus energetics)
of the material inflowing onto galaxies and outflowing from them, temperature
and metallicity gradients throughout the probed halos, etc. 

\stress{To increase the number of IGM hot baryon filaments to thousands and detect $\ge
80$\% of the WHIM mass, to study the physics, kinematics and chemistry of these
baryons in details and so understand the IGM-galaxy co-evolution with cosmic
time,} a much higher resolution soft X-ray spectrometer is needed, with a
resolving power sufficient to resolve thermal (and therefore also turbulent)
velocities of light metals in gas with temperatures of $\sim 10^6$ K. 

\begin{figure}[!htb]
\vspace{-0.5cm}
\includegraphics[width=0.5\columnwidth]{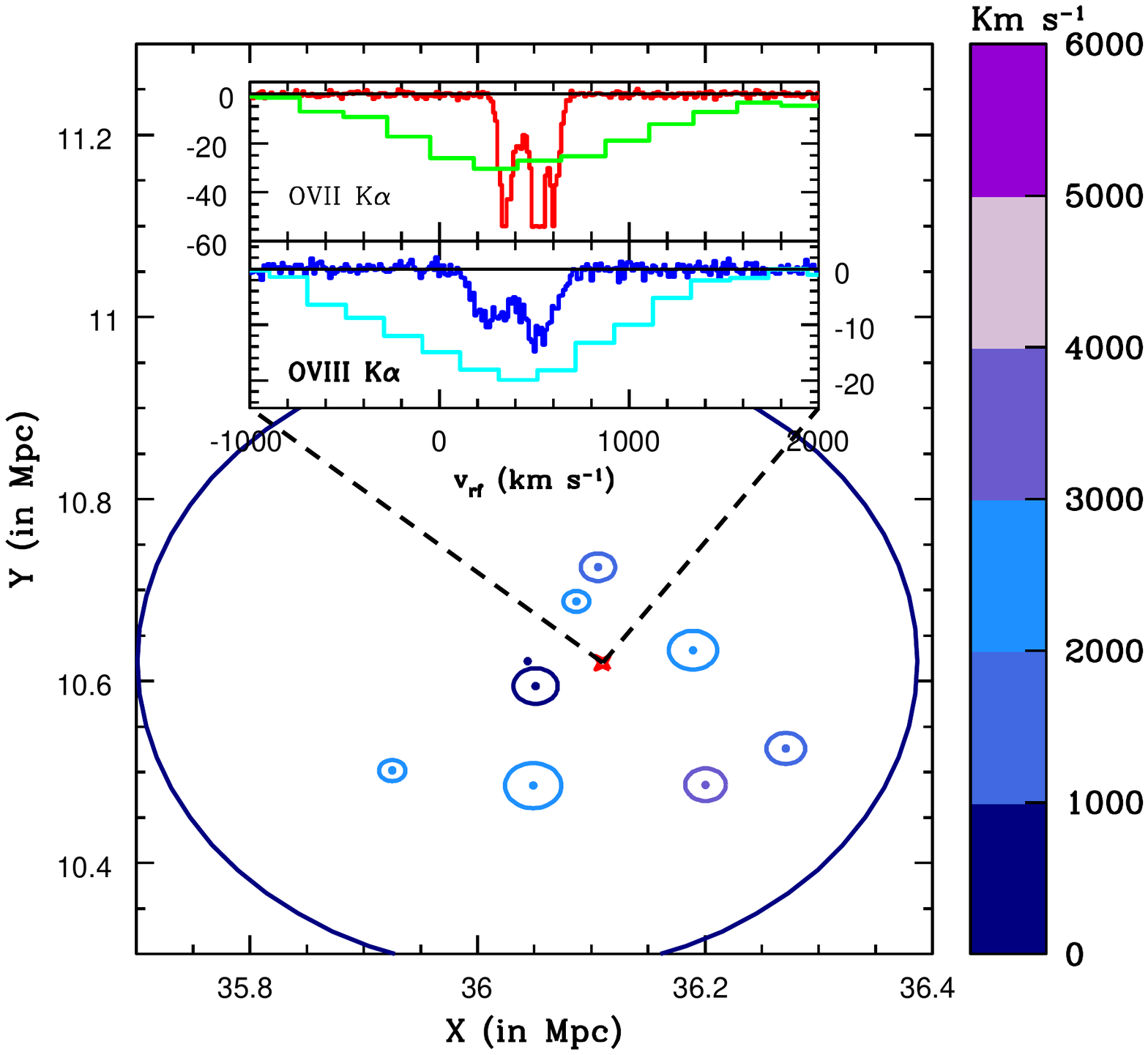}
\includegraphics[width=0.5\columnwidth]{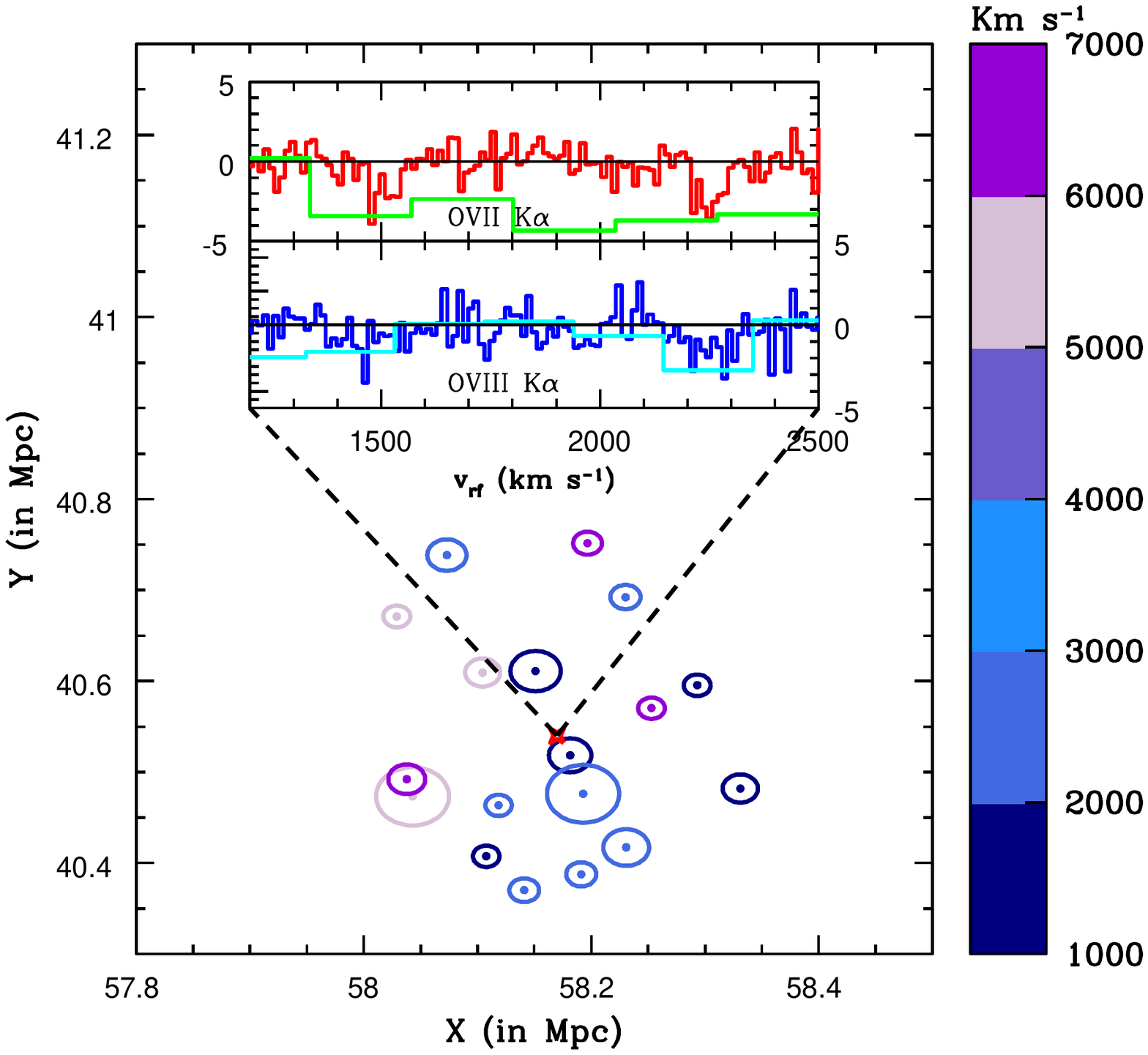}
\vspace{-3.2cm}
\caption{\footnotesize Two distinct regions extracted from the 100 Mpc$^{3}$
periodic box Eagle simulation. The boxes start at $z=0.10064$ (rest-frame
velocity $v_{\rm rf} = 0$) and extend to hubble-flow distances $v_{\rm rf} =
6770$ km\,s$^{-1}$ (the size of the periodic box). Circles are all the halos
with comoving M$_{200} \ge 10^9$ M$_{\odot}$ and within 200 comoving kpc from
the the lines of sight present in the simulations. Circles have radius equal to
half the halo's virial radius in comoving kpc and are color-coded in rest-frame
velocity (i.e. redshift) according to the bar-legend to the right of the
figures. The inserts in the two panels are small portions of mock \hirex (red
and blue histograms) and \athena/\xifu (green and cyan histograms) spectra of
the two lines of sight (red crosses near the center of the boxes), centered on
the \ion{O}{vii} He-$\alpha$ (top panels) and \ion{O}{viii} Lyman-$\alpha$
(bottom panels) transitions at the redshifts of the hot intervening absorbers
present in these two boxes. Mock spectra have been produced with the
fitting package {\sl Sherpa} \citep{Freeman01} adopting a model consisting of
the $\Gamma=2$ power-law continuum of a bright (F$_{0.2-1.35\,{\rm keV}} =
10^{-11}$ erg\,s$^{-1}$\,cm$^{-2}$) background AGN times the opacity to the
metal transition produced by the gas in the simulation boxes, and integrating
for 100~ks (left panel) and 1~Ms (right panel).}
\label{whim}
 \vspace{-0.5cm} 
\end{figure}

Fig. \ref{whim} shows two distinct regions of the 100 Mpc$^{3}$ periodic box
Eagle simulation \citep{Schaye15,Crain15}. These two Eagle regions are
dramatically different in galaxy environment and IGM properties, and have been
chosen to show the vast range of physical, kinematical and chemical conditions
predicted for the local hot IGM depending on the amount of feedback that
different regions of the Universe undergo. The insert of Fig. \ref{whim}a (left
panel) shows \hirex (red and blue histograms) and \athena/\xifu (green and cyan
histograms) mock spectra of two strong and multi-component \ion{O}{vii} (top
panel) and \ion{O}{viii} (bottom panel) absorbers, spanning a range in
rest-frame velocity $v_{\rm rf} \simeq 50-800$ km\,s$^{-1}$. The total EWs are
$49.5^{+2.6}_{-0.6}$ m\AA\ for \ion{O}{vii} and $24.6\pm 0.9$ m\AA\ for
\ion{O}{viii} as measured with \hirex in only 100 ks. These complex absorbers
are clearly associated with the hot gas permeating the largest halo present
along this line of sight (M$_{200c} = 2.8 \times 10^{13}$ M$_{\odot}$), with a
rest-frame (Hubble-flow plus peculiar) velocity $v_{\rm rf}^{\rm Halo} \simeq
470$ km\,s$^{-1}$ and a line of sight impact parameter of only 61 kpc. The
\athena/\xifu will detect both the \ion{O}{vii} and \ion{O}{viii} absorbers, but
only an instrument like \hirex would allow us to \stress{unambiguously resolve their
multi-component nature and precisely measure temperature and density gradients}
across the $\sim 750$~km\,s$^{-1}$ broad \ion{O}{vii} He-$\alpha$ and
\ion{O}{viii} Lyman-$\alpha$ opacity profiles, thus enabling a line-of-sight
tomography of the hot halo medium.

It is worth mentioning that there is also great synergetic potential of \hirex
and imaging X-ray observatories featuring high effective area and equipped with
micro-calorimeters (e.g. \athena) thanks to their capability to detect WHIM in
extended diffuse emission in oxygen lines (boosted by resonant scattering of
the cosmic X-ray background). Combining high-resolution absorption-line studies
by \hirex with such  data will allow the determination of the physical state of
the WHIM gas, e.g. \stress{differentiate truly a diffuse medium of small
overdensity from denser  clumps having a low filling factor} \citep[see,
e.g.,][]{churazov2001,khabibullin2019}.

Fig. \ref{whim} (right panel) shows a dramatically different case. Two much
weaker \ion{O}{vii} and \ion{O}{viii} absorbers are present along this line of
sight, at rest-frame velocities $v_{\rm rf}^1 \simeq 1450-1550$ km\,s$^{-1}$
and $v_{\rm rf}^2 \simeq 2150-2350$ km\,s$^{-1}$. Their EWs, as measured with \hirex
in 1 Ms (red and blue histograms), are for component 1:
$0.50^{+0.11}_{-0.09}$ m\AA\ (5.4$\sigma$, \ion{O}{vii}),
$0.37^{+0.11}_{-0.10}$ m\AA\ (3.7$\sigma$, \ion{O}{viii})
and for component 2:
$0.60^{+0.09}_{-0.09}$ m\AA\ (6.4$\sigma$, \ion{O}{vii}), 
$0.46^{+0.09}_{-0.11}$ m\AA\ (4.2$\sigma$, \ion{O}{viii}). 

\noindent
Absorber 1 is produced by the outskirts (impact parameter of 0.7 virial radii)
of the CGM of a small halo with M$_{200c} = 4.6 \times 10^9$ M$_{\odot}$ and
rest-frame (Hubble-flow plus peculiar) velocity $v_{\rm rf}^{\rm halo} = 1500$
km\,s$^{-1}$. Absorber 2, instead, is not clearly (i.e. with similar $v_{\rm rf}$)
associated with the CGM (or CGM's outskirts) of any halo in the box, and is
therefore imprinted by a truly diffuse intervening WHIM filament with its
external parts hotter than the center (compare the symmetrically broader profile
of the \ion{O}{viii} Lyman-$\alpha$ absorber to the narrower profile of the
\ion{O}{vii} He-$\alpha$ absorber). The \athena/\xifu (green and cyan
histograms) does not resolve the two absorbers and detects only the sum of the
two \ion{O}{vii} (not \ion{O}{viii}) lines, thus not allowing for a clear halo
association. 

This wealth of diagnostics can only be achieved with soft X-ray resolving powers
$R\gtrsim 5000$. 

\subsection{The Circum-Galactic Medium} 

\subsubsection{The CGM of the Milky Way}

Despite many obstacles, the Milky Way is still one of the best laboratories to
test our understanding of galaxy formation and evolution. High resolution X-ray
spectroscopy is essential in addressing several outstanding important questions
in this field. The hot phase of the CGM can only be probed via spectroscopic
features from highly ionised metal species such as \ion{O}{vii} and
\ion{O}{viii}. The soft X-ray band between 0.2--2 keV also offers a unique
opportunity to probe the cold, warm, and hot phases of the Milky Way CGM
simultaneously, and therefore provides a complete census of the baryon content
of the Milky Way. \athena \citep{barcons17} will detect absorption lines with
EWs greater than 5 m\AA\ near the strong lines of \ion{O}{vii} and \ion{O}{viii}
but with poor velocity information. On the other hand, with its high sensitivity
($A_{\rm eff} \sim$1\,500--2\,000~cm$^{-2}$) and high spectral resolution ($R
\sim 10\,000$), \hirex will be able to provide detailed information on the
physical, kinematic, and chemical status of the Milky Way CGM at a level that
cannot be addressed by \athena.

{\center \bf Baryonic content of the Milky Way}
\smallskip

\noindent Recent studies suggested that a large amount of the baryons in the
Milky Way is missing \citep[e.g.,][]{maller04,mcgaugh10}. These missing baryons
are either located in the distant hot halo or dispersed in the local large scale
structure via some feedback processes
\citep[e.g.,][]{fang2013,miller-bregman2015,faerman2017,bregman2018}. However,
the fate of these baryons is very difficult to determine with the current
generation of X-ray telescopes because: (1) the coupling with the Doppler-$b$
(line width) parameter leads to large uncertainties in the column density
measurements of ion species such as \ion{O}{vii} and \ion{O}{viii}; (2) we lack
the information on the ionisation state as well as the chemical abundance of the
CGM. The decoupling of the Doppler-$b$ parameter and the column density can be
achieved by measuring high-order transitions such as the \ion{O}{vii} He$\beta$
line, which so far can only be done with two or three very high S/N sight lines
toward background AGNs. However, \hirex is more than 1\,000 times more sensitive
than the current-generation X-ray spectrometers, and it will allow us to obtain
an accurate measurement of ion column densities for more than 100 AGN sight
lines. Furthermore, with the high spectral resolution, \hirex can clearly detect
and resolve nearly all the ionisation stages of oxygen from \ion{O}{i} to
\ion{O}{viii}, as well as several other ion species. This allows for \stress{a
precise determination of the ionisation mechanism, and the relative abundance of
the metals}. With this information, we expect a complete understanding of the
fate of the missing baryons in our Milky Way galaxy.

\smallskip
{\center \bf Kinematic, physics and chemistry of the Milky Way CGM}
\smallskip

\begin{figure*}[!htb]
\centering
\includegraphics[width=14.truecm]{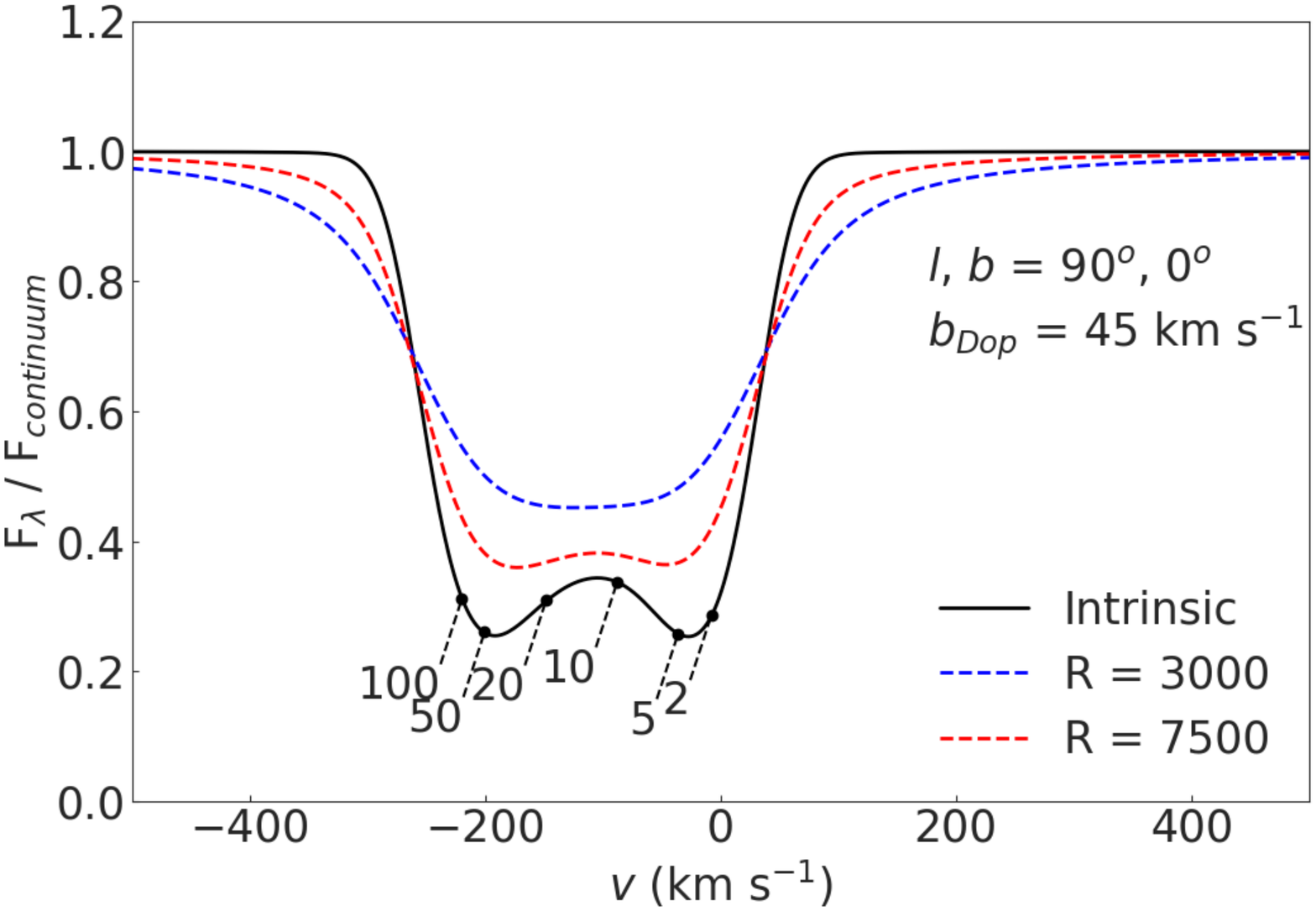}
\vspace{-0.5cm}
\caption{\footnotesize The \ion{O}{vii} He$\alpha$ absorption line from a
rotating Galactic hot halo \citep{hodges16}, modeled with $v_{\phi} = 180$
km\,s$^{-1}$, toward \textit{l} = $90^{\circ}$, \textit{b} = $0^{\circ}$, with
Doppler parameter $b=45$ km\,s$^{-1}$ \citep{miller16a}. The contributions as a
function of distance from the Sun are shown along the bottom of the intrinsic
curve, in kpc. The rotation curve of the hot halo, which is encoded in the line
shape, can be determined from high spectral resolution observations.}
\label{cgmfig}
\end{figure*}

\noindent By providing velocity information, high resolution spectroscopy is
critical in diagnosing the physical, kinematic, and chemical states of the Milky
Way CGM. With velocity information, we can find out \stress{whether the X-ray
absorbing gas is in a state of rotation/inflow/outflow, resolve the absorption
lines into different components, and distinguish between thermal and turbulent
line broadening.} A resolution of $R \sim 5\,000 - 10\,000$ can help match
physically distinct phases and measure bulk velocity better than
$\sim$5--10~km\,s$^{-1}$ (based on the line centroid) and measure the
Doppler-$b$ parameter to better than $\sim$10--20~km\,s$^{-1}$ (based on line
width). Such resolution is necessary to resolve line profiles and so, for
example, discriminate between a rotating and non-rotating hot halo (vital for
understanding the accretion and feedback history of the Galaxy;
Fig.~\ref{cgmfig}), separate different line transitions (e.g., \ion{O}{ii}
K$\beta$ from \ion{O}{vi} K$\alpha$, and \ion{O}{ii} K$\alpha$ from O$_2$
K$\alpha$), resolve bulk velocity for C, O, and Fe, and resolve turbulent from
thermal motion at $T < 10^7$ K. This is a critical advantage of \hirex over
\athena as the latter, with a resolution of $R \sim 200$, can detect lines
without being able to resolve them.

\subsubsection{The CGM of external galaxies}

\athena will likely detect extended hot halo gas around $L_\star$ \citep[or Schechter Luminosity,][]{Schech76} 
galaxies to about 100 kpc \citep[according to models;][]{breg15} but with poor velocity
information. These hot halos are likely to extend far beyond 100 kpc (250--500
kpc; Fig. \ref{whim}), based on models \citep[e.g.,][]{Schaye15} as well as the
observations of stacked Sunyaev-Zeldovich measurements \citep[e.g.,][]{Ma2015}.
The extent of the hot gas, as well as its mass, composition, and temperature are
set by the accretion and feedback processes, which are poorly constrained. These
fundamental properties will be determined by the \hirex spectroscopic mission,
providing great advances in our understanding of galaxy formation and evolution.

{\center \bf Missing metals and baryons in external galaxies}
\smallskip

\noindent The optical parts of galaxies and their gaseous disks are missing
about 75\% of the metals produced by their stars and 70-90\% of the initial
baryon content \citep{peep14,mcgaugh10}. The inner 100 kpc around galaxies will
be probed with \athena, but the gas and metal masses increase with radius, so
most of the baryons and metals lie beyond 100 kpc. These outer regions will be
studied with \hirex, which has a sensitivity about an order of magnitude better
than \athena for line detections. It will measure absorption in more than 100
galaxy halos, providing an excellent radial profile of ions (probably to
250--500 kpc, 1--2R$_{200}$) in the long-lived halo where the cooling time
exceeds a Hubble time. The \ion{O}{vii} columns can be estimated from an
extrapolation of Milky Way gas and from direct detection along one extragalactic
sight line \citep{breg15,Nicastro18}. These \ion{O}{vii} columns are more than
an order of magnitude greater than observed UV \ion{O}{vi} sight lines
\citep{Savage2003,Werk2016} from cooler gas (also true for lower ionisation
state lines), so these X-ray absorption lines will yield \stress{a definitive measure of
the metals, determining whether the missing metals lie within an extended hot
halo.} This relates directly to the formation and evolution of galaxies.

Single lines of sight will detect only a few lines (He-like and H-like C and O),
but by stacking the absorption systems, weaker lines will be revealed. These
include higher ionisation lines than \ion{O}{vii}, such as \ion{O}{viii},
\ion{Ne}{ix}, and \ion{Ne}{x}, additional lines at the typical ambient
temperature of the halo, such as \ion{N}{vi}, \ion{N}{vii}, as well as lower
ionisation state lines, such as  \ion{O}{i}, \ion{O}{iii}, \ion{O}{iv}, and
\ion{O}{v}. These lines will define the temperature distribution of ions, which
is caused by cooling and feedback processes and has become an important
prediction of models.

{\center \bf Feedback and rotation in external galaxies}
\smallskip

\noindent Feedback from SNe and AGNs regulates the formation of galaxies and
their environment. Detailed measurements of the hot X-ray gas provide a snapshot
of the feedback today. Feedback will cause the outflow of gas from the disk and
there will be turbulence caused by these motions \citep[e.g.,][]{soko2018}. The
net effect will be to broaden the lines beyond the thermal widths (40--50
km\,s$^{-1}$), and \hirex will have the velocity resolution to separate a line
into its thermal and turbulent components, the latter being a direct measure of
feedback.

A nascent galaxy gains angular momentum from surrounding non-uniform regions,
with the rotational velocity increasing as the gas flows to smaller radii
\citep{Lagos2017}. This leads us to expect the hot halo to be rotating, although
the initial value of the specific angular momentum can be decreased by the
accretion of other halos, and if feedback is very effective, it will mix the
angular momentum of the hot halo gas. Thus the rotational profile of the hot
halo contains vital clues about accretion and feedback. \hirex will measure the
velocity of line centers to exceptional accuracy ($<$30~km\,s$^{-1}$) for sight
lines through many different galaxy halos, allowing one \stress{to assemble a
composite rotation curve to several hundred kpc.}

\subsection{Outskirts of clusters and groups of galaxies\label{sect:outskirts}}

 \begin{figure*}[!htb]
 \centering
 \includegraphics[width=14.truecm]{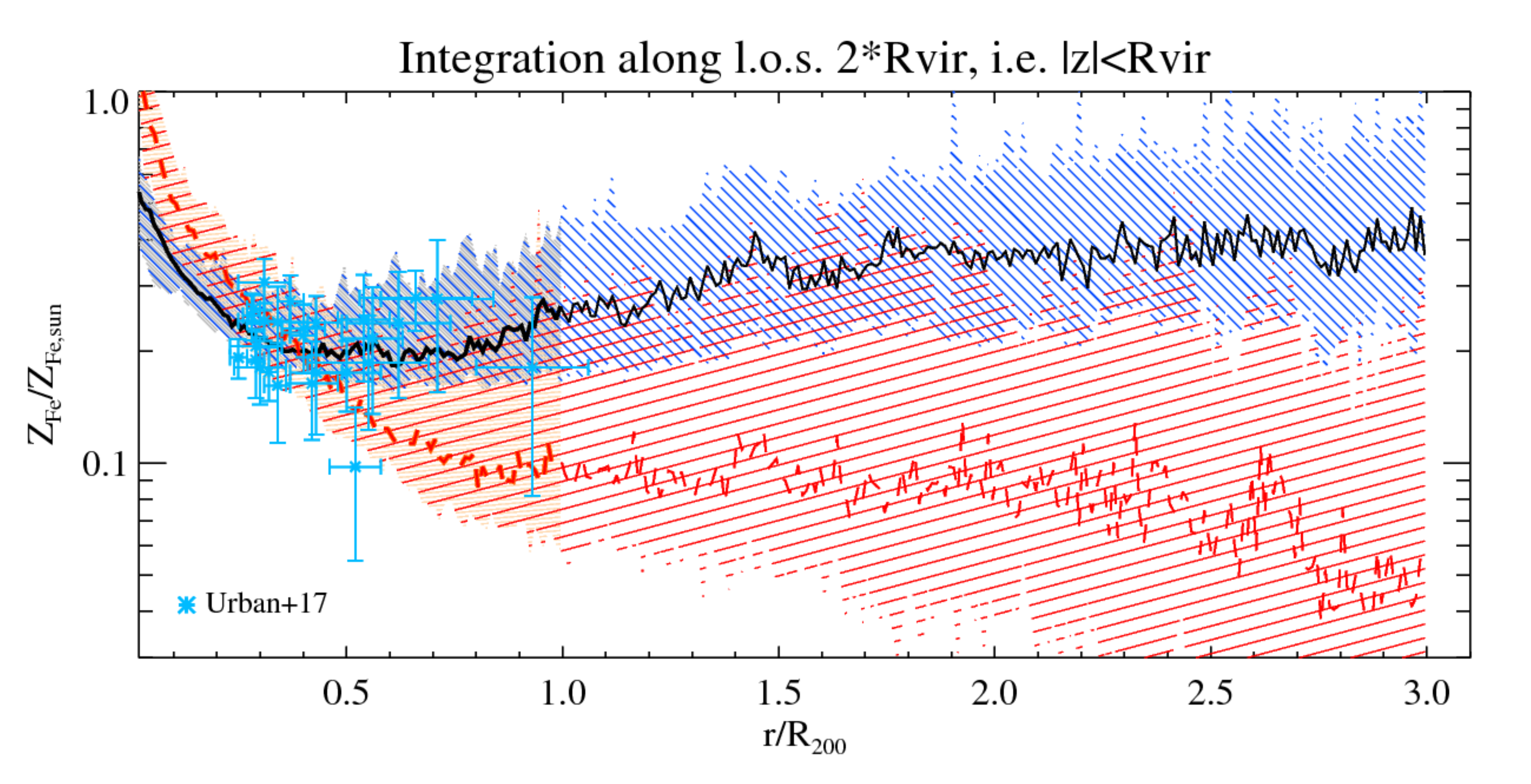}
 \caption{\footnotesize Comparison between observed metallicity profiles in the
 outskirts of galaxy clusters and results from cosmological
 hydrodynamic simulations. The cyan asterisks with errorbars
 correspond to the observational results by \protect\cite{Urban17}
 for 10 nearby clusters observed with \suzaku. Simulation
 results corresponds to the set of simulated clusters whose
 metallicity profiles have been studied in \protect\cite{Biffi18},
 with blue and red corresponding to simulations that include and
 exclude the effect of AGN feedback, respectively. The two curves
 correspond to the median profile computed within the simulation
 set, while the shaded areas mark the corresponding
 r.m.s. scatter.}
 \label{metprof}
 \end{figure*}
 
 Clusters and groups of galaxies represent unique signposts in the Universe
where both thermo- and chemo-dynamical properties of the diffuse ionised plasma
can be studied in detail through X-ray observations. The past (\asca and
\sax) and the current generation (\xmm, \chandra and \suzaku) of
X-ray satellites have demonstrated that metals are not homogeneously distributed
in the intra-cluster medium (ICM; see \cite{Biffi18b, Mernier18} for recent
reviews on results from simulations and observations). Indeed, relaxed cool-core
clusters are characterised by negative metallicity gradients, with an enhanced
metallicity in the core regions, likely to be associated to the process of star
formation of the galaxies whose assembly gave rise to the Brightest Cluster
Galaxies (BCGs) that are at the centre of these clusters. On the other hand,
dynamically disturbed non--cool core clusters have a much flatter metallicity
gradients, with a lower level of enrichment in central regions. The reason for
this is due to the mixing of the gas due to the same dynamical processes which
lead to the disruption of the cool core. While these observational results hold
out to about one quarter of the virial radius, $R_{\rm vir}$, the situation is less
clear at larger radii, where reliable measurements of metallicity are hampered
by the low level of emissivity and the corresponding need for precisely
characterising background contamination. Indeed, the low particle background of
the \suzaku satellite recently allowed to trace, although with rather large
uncertainties, the ICM metallicity out to scales approaching the virial radius
of a handful of nearby clusters (see Fig.~\ref{metprof}).

These studies gave a clear indication of the inextricable link between the
recent dynamical history of galaxy clusters and the pattern of chemical
enrichment within regions of galaxy clusters and groups which encompass a
fraction of their virial region.

Quite interestingly, cosmological hydrodynamic simulations that include a
detailed description of chemical enrichment converge to indicate that a negative
metallicity gradient in the central regions of galaxy clusters and groups
naturally arises as the result of the processes of star formation and
hierarchical assembly of such structures, independent of the efficiency and
nature of the sources of energy feedback that should regulate star formation. On
the other hand, the same simulations also predicted that largely different
patterns of chemical enrichment are expected in the outer regions of galaxy
clusters when including different feedback sources: while feedback from
supernovae generates a negative metallicity gradient extending out to the virial
radius and beyond, the inclusion of AGN feedback causes a flattening of the
metallicity profiles beyond about half of $R_{\rm vir}$
\citep[e.g.,][]{Fabjan10,McCarthy11}. In fact, this different behaviour is due
to the efficiency with which AGN feedback expels metal-enriched gas from the CGM
of the proto-cluster galaxies at redshift $z\simeq 2$--4, when gas accretion
onto SMBHs reaches the peak of activity. This causes in turn a pre-enrichment of
the IGM, which leads to a sort of metallicity floor at low redshift
\citep[e.g.,][]{Biffi18}. This effect is illustrated in Fig. \ref{metprof},
which shows the metallicity profiles predicted by simulations both including
(red) and excluding (blue) the effect of AGN feedback, and compares them to
observational data on ICM metal enrichment out to the largest radii sampled so
far. 

While the much higher sensitivity and spectral resolution of \athena/\xifu are
expected to significantly improve with respect to the current observations out
to $R_{500}$ \citep{Cucchetti18}, mapping the metallicity of the ICM in emission
beyond $R_{\rm vir}$ could become prohibitively expensive, or even hardly
possible, due to the requirements on the background control and collecting area.

At the same time, while current observations of cluster outskirts mainly trace
Fe (by mainly measuring the equivalent width of lines in the Fe-K complex),
tracing different chemical species would provide further insights on the
processes of feedback and gas circulation that determine the cosmic cycle of
baryons in general. For instance, Fe and O are expected to be produced by
different stellar populations, which are characterised by different lifetimes
and, therefore, release them over different time-scales. As a consequence, their
distribution could differ, depending on the efficiency of the dynamical
processes (e.g. outflow, turbulence) driving their diffusion
\citep{Simionescu19}.

\begin{figure*}[!htb]
\centering
\vspace{-1.5cm}
\includegraphics[width=10.truecm]{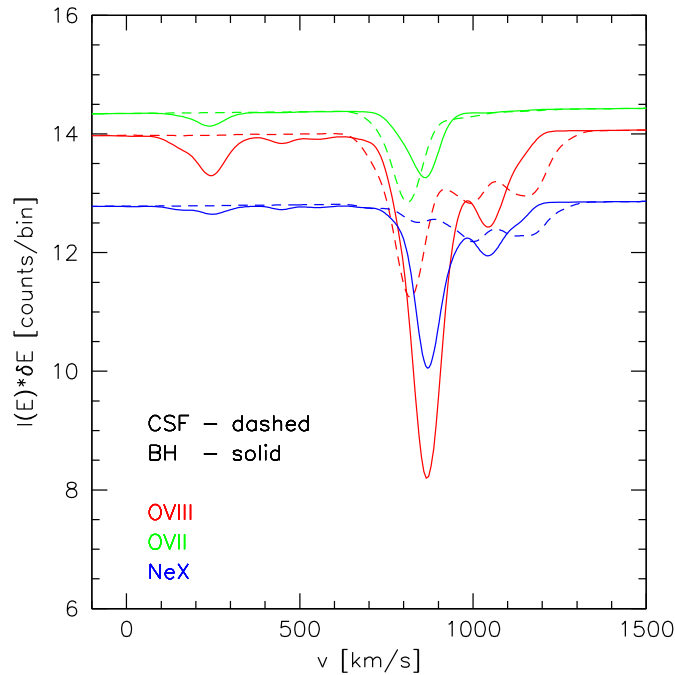}
\vspace{-2.5cm}
\caption{\footnotesize Velocity profile of the absorption lines produced by
\ion{O}{vii}, \ion{O}{viii} and \ion{Ne}{x} ions in the spectrum of a background
AGN at a projected distance of $\sim$1.6~Mpc from the center of a massive
cluster from simulations of \citet{Biffi18}. Compared to WHIM, the role of
\ion{O}{viii} and \ion{Ne}{x} is enhanced due to higher temperature of the gas.
The dashed and solid lines correspond to the CSF and BH runs, correspondingly.
Two prominent absorption structures are present for this sight-line, one at
+900~km\,s$^{-1}$ and a fainter one at +300~km\,s$^{-1}$. Each structure
possesses an additional kinematic substructure on top of the thermal broadening.
As expected, the lines are significantly stronger in the BH run. The vertical
axis shows the number of counts per resolution element for the AGN with 0.5--2
keV flux $\sim 2\,10^{-13}$\,erg\,s$^{-1}$\,cm$^{-2}$ and photon index
$\Gamma=1.2$ for a 1\,Ms observation with 1500~cm$^2$ effective area.}
\label{fig:grating-s}
\end{figure*}

Fig.~\ref{fig:grating-s} compares the \ion{O}{vii}, \ion{O}{viii} and 
\ion{Ne}{x} absorbing systems predicted by  the cooling-star formation-supernova
feedback (CSF) and AGN feedback (BH) simulations \citep{Biffi18}. As expected,
in the BH simulation that involves redistribution of metals by AGN feedback over
cluster  outskirts, the absorption lines are much more prominent. The lines are 
resolved and can be used to study the flow of metals in cluster  periphery.
Note, that the higher temperature than in the WHIM leads to  stronger lines of
\ion{O}{viii} and \ion{Ne}{x}. This property can be used to separate  the WHIM
contribution from the hotter cluster outskirts.

High spectral resolution (10--50 times better than calorimeters) absorption
tomographic studies of cluster gas allow us \stress{to trace different elements
at different ionisation stages, and to measure  their spatial distribution and
motion along the line-of-sight at large radii} (out to $2R_{\rm vir}$) that are
impossible to do in emission. 

\subsection{AGN and galactic outflows}

Current X-ray observatories have established the presence of outflows, produced
in the vicinity of supermassive black holes (SMBH) at the heart of AGN. The
so-called warm absorbers are detected in a large number of AGN via a wealth of
soft X-ray absorption lines from C, N, O, Ne, Mg, Si and S, tracing low
ionisation gas outflowing at 100s to 1000s of km\,s$^{-1}$ and located at a few
10s--100s of parsecs from the SMBH \citep{Kaastra2000,Kaspi2002,Blustin2005}. In
obscured Seyfert galaxies, this medium is observed via the presence of numerous
soft X-ray emission lines. The temperature, density and the source of the
ionisation equilibrium of the emitting gas can be measured from radiative
recombination continua (RRC) and He-like triplets, as well as from a full
characterisation of satellite lines and the Fe L `forest'
\citep{Liedahl1999,Porquet2010}. 

There is increasing evidence for the presence of blue-shifted Fe K-shell
absorption lines in AGN at rest-frame energies higher than 7 keV
\citep{Tombesi2010,Gofford2013}. These so called ultra fast outflows ("UFOs")
are likely driven off the accretion disk by either radiation pressure
\citep{Proga2000} or magneto-rotational forces \citep{Kato2004,fukumura2015}, or
both, and the outflow rates derived can be large, of order several solar masses
per year. Furthermore these fast, energetic winds, may be the initial stage of
the sweeping process that leads to mass losses of hundreds to thousands of
M$_{\odot}$\,yr$^{-1}$ on the scale of the AGN host galaxy
\citep{Tombesi2015,Feruglio2015}. The presence of soft X-ray absorption lines
from UFOs has been detected too, showing that these fast winds exhibit different
ionisation states
\citep{Longinotti2015,Reeves2018,Danehkar2018,Pinto2018,Serafinelli2019}. These
arise from He and H-like O, Ne and L-shell Fe lines and are most apparent when the
spectrum is more absorbed overall. This may be associated with lower ionisation
gas, like the Broad Absorption Line (BAL) winds in the UV, as part of a clumpy
phase of the disk wind capable of producing substantial X-ray obscuration
\citep[e.g., NGC\,3783,][]{Mehdipour2017}. This process can then facilitate
radiative acceleration and driving of winds in powerful quasars at higher
redshifts, which are currently beyond reach for soft X-ray spectroscopy.

It is essential to measure all these wind phases over a wide range of radial
distances and velocities to understand their nature, their link, their launching
mechanism(s) and duty cycle, and thus, ultimately, their relevance in polluting
the host galaxy's surroundings with metals, as predcited by simulations (e.g.
Sect.~\ref{sect:igmz2}--\ref{sect:outskirts}, Fig.~\ref{metprof}). This can
only be performed via accurate measurements of the wind physical properties
(e.g., column density, density, ionisation parameter, velocity, distance, time
variability) and their relationships, which, in turn, requires high spectral
resolution and throughput in the soft X-rays. High resolving power
R$\sim$5\,000--10\,000 is also critical for an \stress{accurate determination of
the total wind kinetic power required for galaxy feedback}. The X-ray wind
energetics can then be compared to that of larger scale molecular and ionised
outflows to assess their efficiency in driving (via either energy or momentum)
feedback on galaxy scales \citep{Fiore2017}. A soft X-ray grating with the
parameters of \hirex (Sect.~\ref{sect:concept}) is also required to robustly
measure wind variability in order to deduce its characteristics.

\begin{figure}[!htb]
\includegraphics[width=0.5\columnwidth]{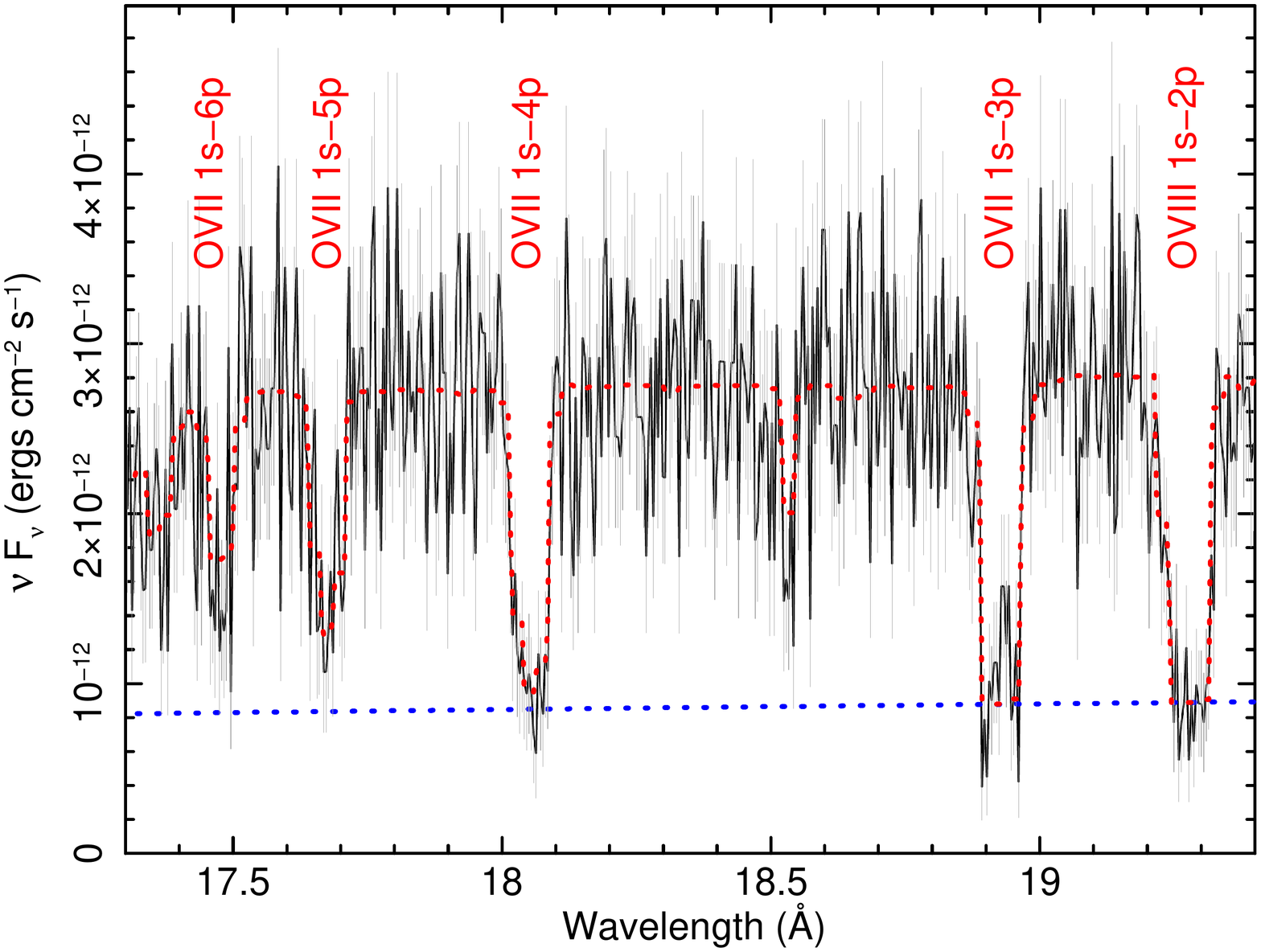}
 \includegraphics[width=0.5\columnwidth]{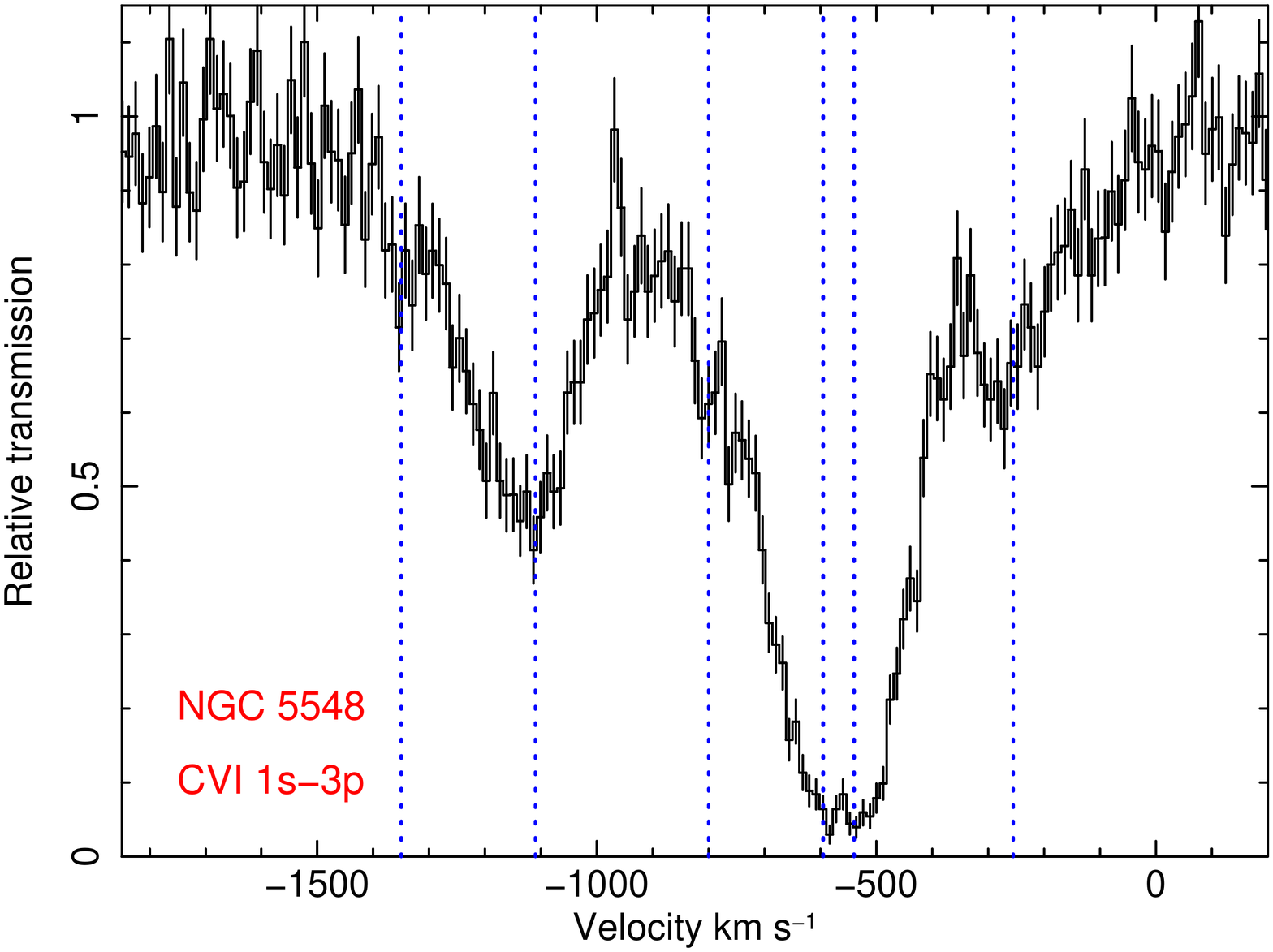}
 \caption{\footnotesize
Left -- simulations of a 20\,ks time slice of the variable, fast outflow in PG
1211+143, with a resolution of $R=5000$. The higher order absorption
\ion{O}{vii} line series is seen in PG\,1211+143, where the strongest saturated
lines reveal the covering fraction and the higher order lines determine the
column density. Right -- a 120\,ks simulation of the Seyfert 1 galaxy, NGC\,5548,
at a resolution of $R=10000$, showing the wind profile in C\,\textsc{vi}. The
wind can be resolved into at least five distinct velocity components (vertical
dashed lines), enabling the wind properties, such as its ionisation, to be
mapped as a function of the outflow velocity. A turbulent velocity of
$\sigma=70$\,km\,s$^{-1}$ was adopted.}
\label{Fig:Fig_AGN}
 \vspace{-0.2cm} 
\end{figure}

The nearby QSO, PG\,1211+143 (at $z=0.0809$) has one of the prototype ultra fast
outflows at soft X-rays. The current \xmm \rgs and \chandra \hetgs grating
observations have a combined total exposure exceeding 1\,Ms
\citep{Reeves2018,Danehkar2018}. However the line diagnostics are currently
limited to the strongest $1s\rightarrow 2p$ lines, due to the low effective area
and resolution, while the mechanism for the outflow variability is uncertain.
High throughput and resolution at soft X-rays are prerequisite for measuring
absorber variability. Fig.\,\ref{Fig:Fig_AGN} (left panel) shows a \hirex
simulation of a short 20\,ks (typical variability timescale of AGN in the
X-rays)  time-slice of PG\,1211+143, focused on the \ion{O}{vii} line series,
demonstrating that the wind variability  can be accurately measured on these
short timescales. The strongest line profiles (e.g. \ion{O}{viii} $1s\rightarrow
2p$ and \ion{O}{vii} $1s\rightarrow 3p$) are saturated, providing a direct
measure of the gas covering fraction, to within $60\pm5$\%. Variations in
covering could correspond to the transverse motion of outflowing clouds across
the line of sight on dynamical size-scales of a few gravitational radii. The
weaker higher order absorption lines are not saturated and yield an accurate
measure of the hydrogen equivalent column density to within $N_{\rm
H}=2.9\pm0.2\times10^{21}$\,cm$^{-2}$. Changes in ionisation in response to a
decrease in the X-ray continuum, via recombination, can be also be measured to
an accuracy of $\pm5$\% on a 20\,ks timescale. Such monitoring enables
\stress{the density, location and size scale of the absorber} to be directly
measured \citep[e.g.,][]{Nicastro99}. The data would also allow us to determine
the outflow properties, such as the column, covering and ionisation, as a
function of the wind velocity. The wind energetics can then be robustly assessed.

The second example is of the Seyfert 1 galaxy, NGC\,5548 (at $z=0.017175$). This
AGN has a notable outflow, spanning at least an order of magnitude in outflow
velocity and two orders in ionisation \citep{kaastra2014}. Presently these
outflow components can only be resolved in velocity space through high
resolution UV spectroscopy, but not in X-rays; e.g. neither at the resolution of
the current X-ray gratings on-board \xmm or \chandra, nor, in ten years from
now, with \athena below 1\,keV. Fig.~\ref{Fig:Fig_AGN} (right) shows a 120\,ks
simulation of the outflow in NGC\,5548, based on its un-obscured X-ray state.
For illustration, the velocity profile at the \ion{C}{vi} $1s\rightarrow3p$ line
is plotted, where a resolution of $R=10000$ will fully resolve all the wind
components. In this example at least five distinct velocity components are
resolved, with $v=250, 550, 800, 1100, 1350$\,km\,s$^{-1}$, each measured to a
typical accuracy of $\pm10$\,km\,s$^{-1}$. Furthermore the wind may vary in
ionisation as a function of velocity, e.g. as can occur if the highest
ionisation gas originates from the fastest wind components. The high resolution
X-ray spectrum would make it possible \stress{to measure all the wind properties
well beyond those measurable in the narrow ionization windows offered by UV
spectroscopy, such as ionisation, as a function of the outflow velocity,
providing a multi-dimensional map of the wind and ultimately revealing its
complete physical structure}.

\section{The Inter-Stellar Medium}

\subsection{Chemical composition and dynamical structure of the ISM}

The diffuse interstellar medium (ISM) has a crucial role in the evolution of the
Galaxy. Different elements are produced by various types of stellar phenomena
(e.g. supernovae type Ia, core-collapse and AGB winds) and their abundances are
the direct testimony of the history of stellar evolution. In the neutral phase,
heavy elements like iron, calcium, magnesium, etc. are mostly locked into solids
and depleted from the gas phase. However, the interstellar dust composition is
not well known and the total (gas + molecules + dust) abundances are yet to be
accurately determined. At long wavelengths, for instance, it is still
challenging to distinguish between carbonaceous and silicate dust grains
\citep[e.g.,][for the $10\mu$m feature]{Min07}.

The soft X-ray energy band ($\sim$\,0.2--2\,keV) contains the strongest
transitions of the K shell from the most abundant interstellar atomic species
(C, N, O, Ne, Mg, Si) and the Fe L shell. Dust and molecules also imprint a
forest of spectral features in the form of X-ray absorption fine structures
(XAFS) supporting multi-wavelength studies of interstellar phases and synergies.
It is possible to probe the ISM complex composition through the study of
interstellar absorption features in the high-resolution X-ray spectra of
background galactic X-ray binaries and active galactic nuclei.

High-quality grating spectra unambiguously show the presence of interstellar
absorption  lines and edges in every spectrum. The whole series of oxygen
(\ion{O}{i-viii}), \ion{N}{i-ii}, \ion{Fe}{i} L$_2$ and L$_3$ edges,
\ion{Ne}{i-x} and many more species are found
\citep[e.g.,][]{Juett04,Juett06,Yao09}. Strong evidence of dust and molecules in
the form of complex XAFS was found in the absorption edges of elements like O,
Mg, Si and Fe that are significantly depleted  from the gaseous phase
\citep[e.g.,][]{Lee05,Lee09,deVries09}. Between 15--25\% and 65--90\% of the
total amount of \ion{O}{i} and \ion{Fe}{i} (and possibly higher fractions for
\ion{Mg}{i} and \ion{Si}{i}) could be locked in dust grains
\citep{Pinto10,Pinto13,Zeegers17,Rogantini18}. Solids may therefore be a
significant reservoir of metals and provide a solution to the problem of the
missing oxygen, but the uncertainties in the atomic databases and the limited
spectral resolution of current detectors ($R\sim$300--500 for \rgs and \letgs)
undermine the detection and measurements of dust features \citep[see,
e.g.,][]{Garcia05,Gatuzz13}. A next-generation detector with superior effective
area ($A>1000$\,cm$^{2}$) and spectral resolution ($R>1000$) at $E\lesssim
1.5$~keV is necessary \stress{to resolve each individual absorption line from
the dominant ions and dust compounds.}

Micro-calorimeters, like \athena/\xifu and \xrism\resolve, will have $R<200$ at
$E=0.5$~keV and thus cannot serve to this particular scope.
Wavelength-dispersive grating spectrometers are needed. Here we adopt a
grating-spectrometer (\hirex) with $R=5000$ resolving power  and 2000\,cm$^{-2}$
effective area at the Fe L and O K edges \citep[e.g.,][]{Heilmann17}. 

\begin{figure}[!htb]
 \includegraphics[width=0.475\columnwidth]{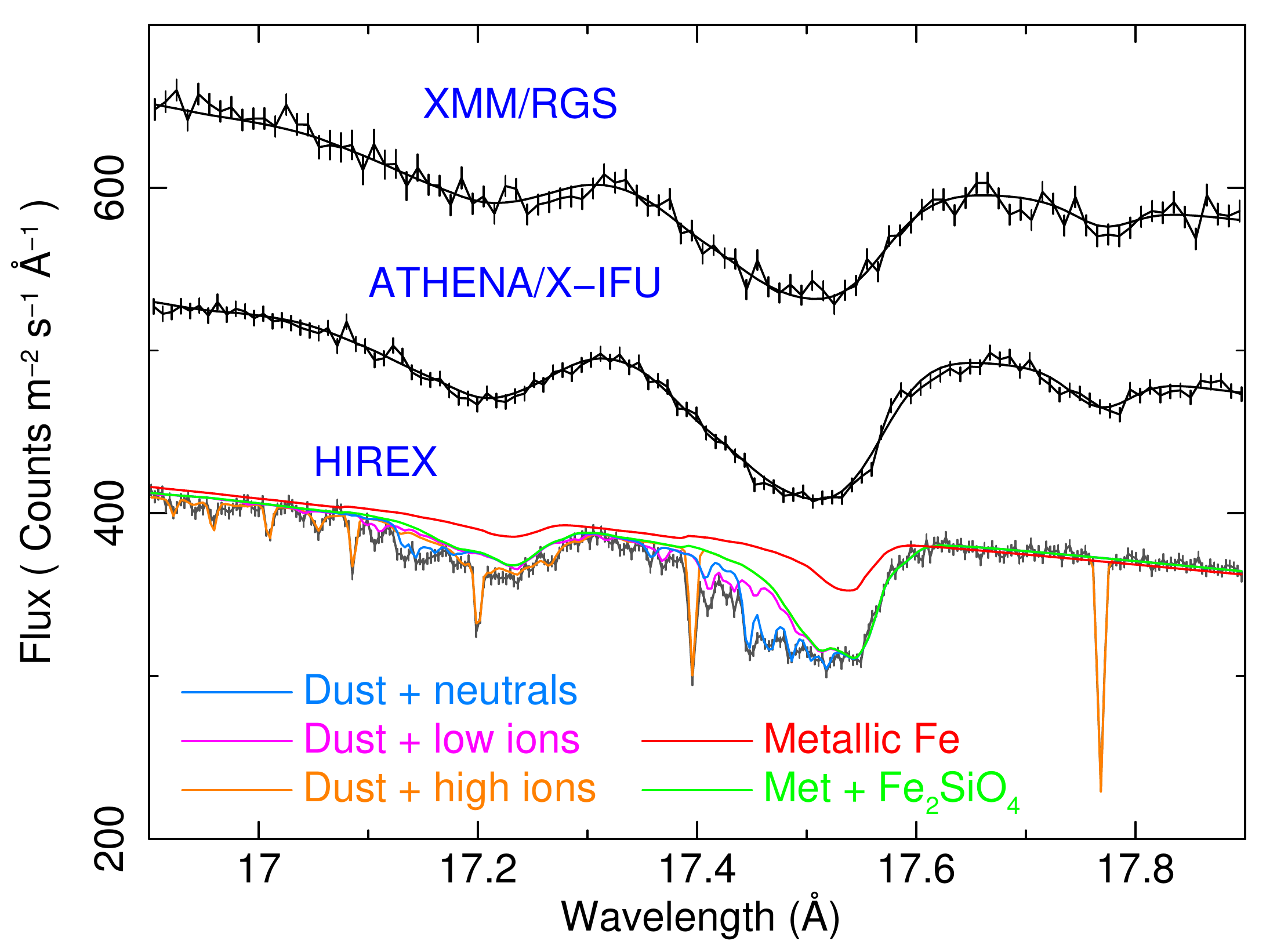}
 \includegraphics[width=0.525\columnwidth]{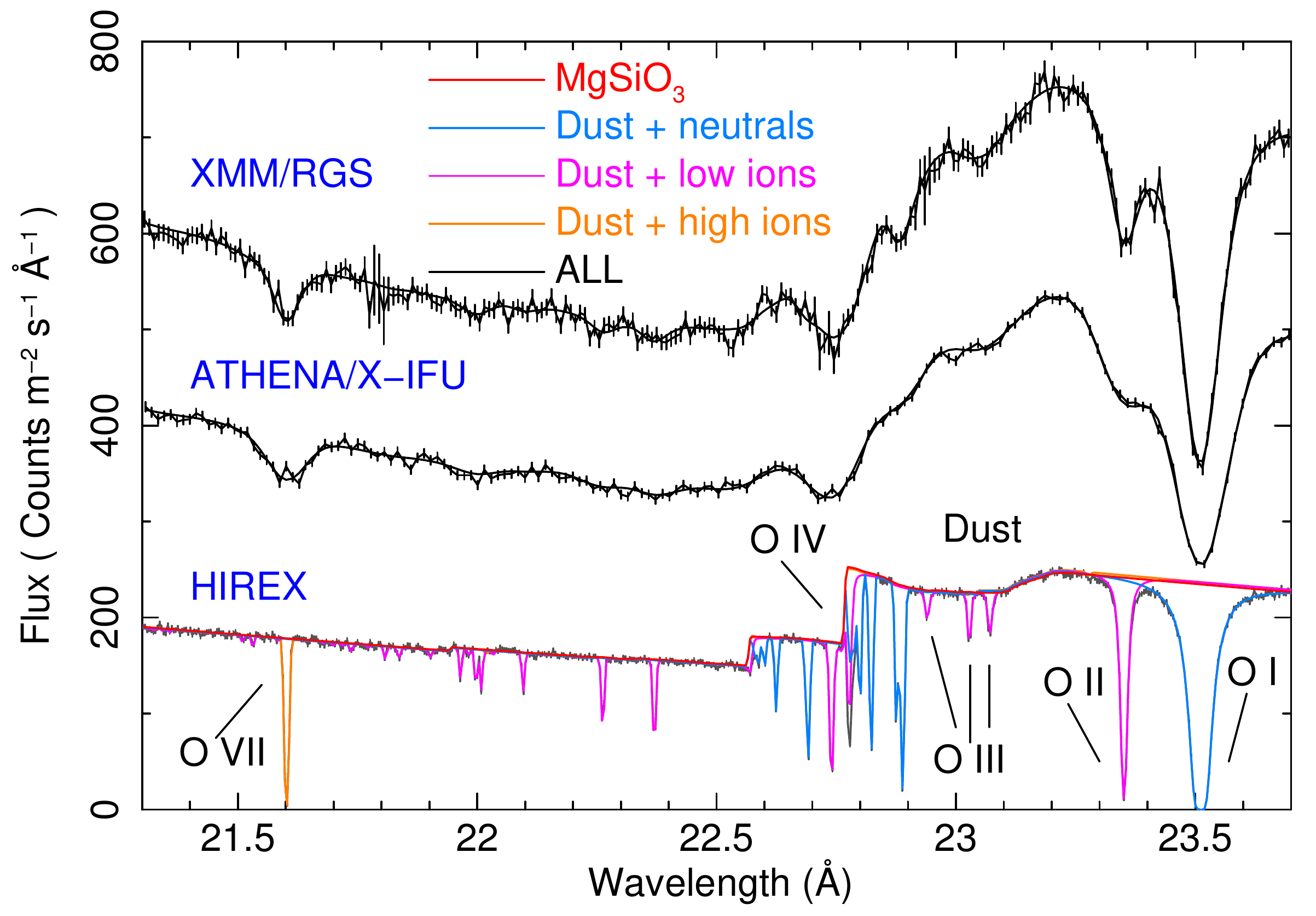}
 \includegraphics[width=0.5\columnwidth]{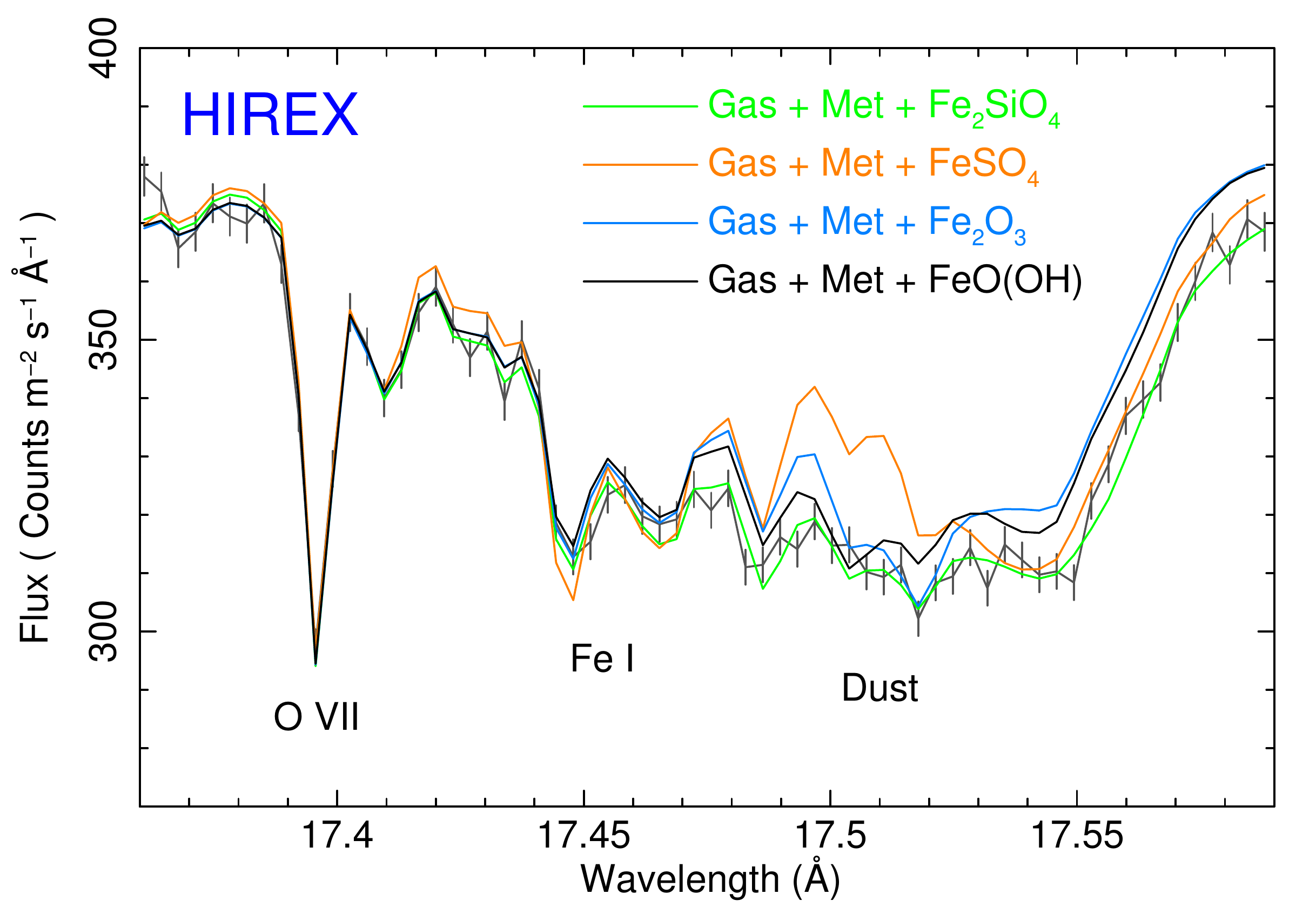}
 \includegraphics[width=0.5\columnwidth]{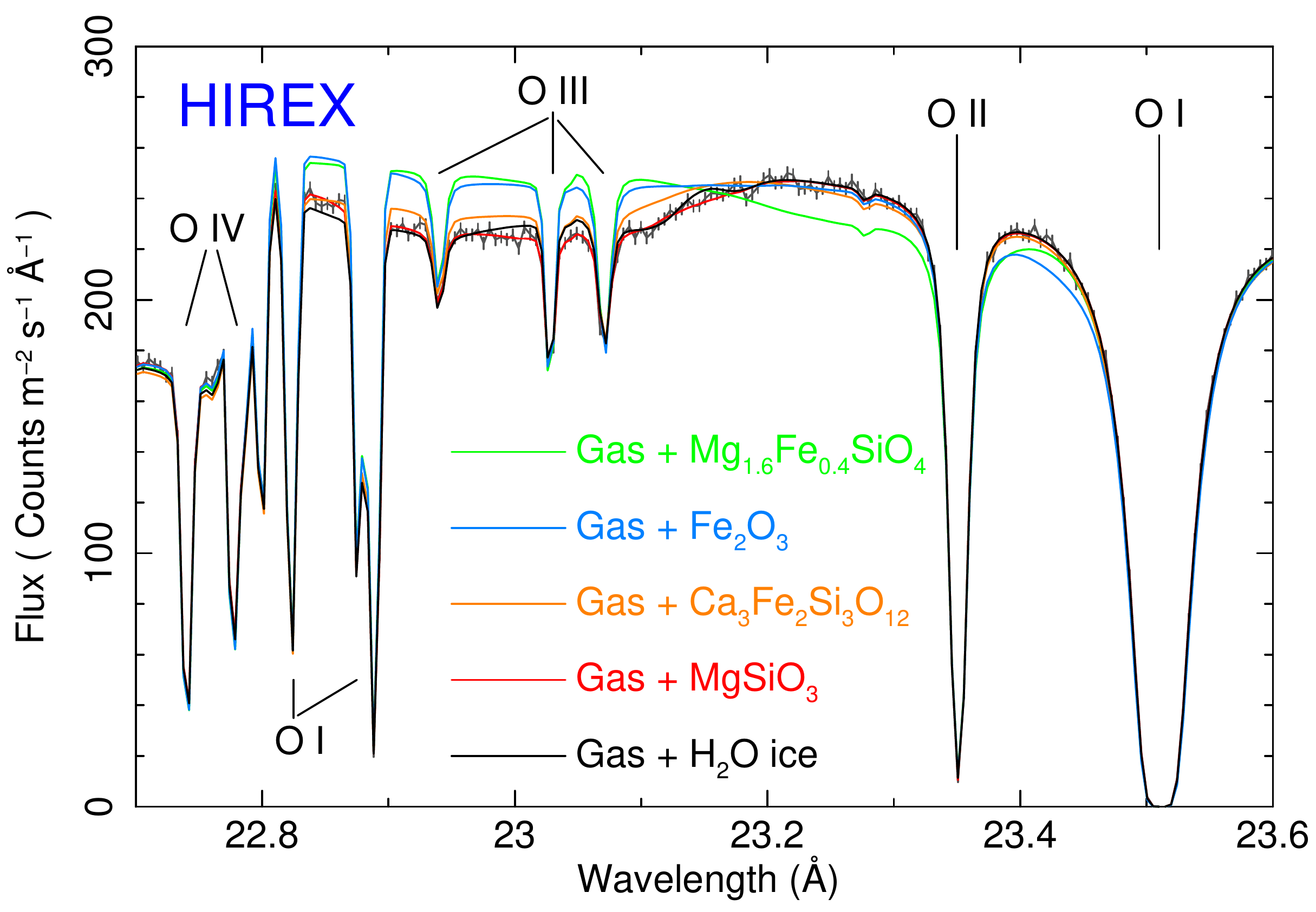}
 \caption{\footnotesize Cygnus X-2 \hirex 50\,ks spectrum simulated using the \rgs fit model
 as compared to \athena/\xifu and \xmm/\rgs
 (top panel), which are shifted along the Y-axis for displaying
 purposes.
 Low ions refer to e.g. \ion{O}{i-v} and high ions to \ion{O}{vi-viii}.
 Only \hirex will be able to resolve individual lines.} 
 \label{Fig:Fig_ISM}
 \vspace{0.cm} 
\end{figure}

We use as template the best fit model for the \rgs stacked spectrum of Cygnus
X-2 which includes a comprehensive description of the ISM (neutral and ionised
gas plus dust). The dust mainly consists of pyroxene (MgSiO$_3$), metallic iron
and fayalite (Fe$_2$SiO$_4$). We have simulated a Cygnus X-2 high-resolution
spectrum with an exposure time of 50~ks (see Fig.\,\ref{Fig:Fig_ISM}).

All the relevant absorption features are well resolved, even the
\ion{O}{iii} doublet at 23.05--23.1\,{\AA}. The accuracy on the column
densities of each molecular compound would be around few\,\% or better. The
uncertainties on the velocities of each gas component will be less than few
km\,s$^{-1}$, providing for the first time \stress{accurate measurements on both line
widths and line of sight velocities}. With the current detectors we can only get 
upper limits of about 100--200 km\,s$^{-1}$. Neither \rgs nor \athena\xifu are
able to resolve and distinguish the individual features. The uncertainties on
gas column densities and temperatures will be better than 5\%, enough \stress{to
distinguish between photoionisation and collisional equilibrium}.

Similar studies will be possible on the ISM of nearby galaxies, by using local
X-ray binaries or Ultra-Luminous X-ray sources (ULXs) as beacons, and on the ISM
of distant galaxies by observing GRB X-ray afterglows soon after the prompt
emission and during their fading phase. 

\section{Metals in-and-around stars}

\subsection{Stars}

To constrain the metal processing operated by stars  the understanding  of
stellar formation and evolution, for both high- and low-mass stars, needs to be
improved. 

High-mass stars are crucial for a wide range of astrophysical aspects: the
starburst events, the chemical enrichment of the Universe, and the
multi-messenger astronomy via gravitational wave (GW) events. These stars have a
strong impact on their Galactic environment through their winds and explosions
as supernovae. When in binaries, they are the progenitors of double-compact
systems which finally merge, emitting GWs. Soft X-rays in massive stars arise in
shocks linked to their winds \citep[e.g.][]{GuedelNaze2009}. X-ray line
morphologies are sensitive probes of the properties of  these winds
\citep[e.g.][]{HerveRauw2013}, providing crucial information on many aspects of
these outflows.  The limited sensitivity of existing facilities and the limited
spectral resolution in the soft X-ray domain of forthcoming calorimeters are
insufficient to resolve these lines and to study their temporal evolution.  A
high-resolution and high-efficiency spectrometer in soft X-rays will not only
help constrain the mass-loss rates of massive stars, but will allow
\stress{tracing the impact of small and large-scale wind structures}. Doppler
tomography of resolved X-ray lines will \stress{map the plasma in the
co-rotating magnetosphere} of magnetic massive stars. The proposed mission will
further allow studying the propagation of photospheric pulsations throughout
the stellar wind, investigating the formation of radiative recombination
continua, and constraining the fraction of the stellar wind that is under the
form of plasma.

\begin{figure}[!htb]
\begin{center}
\includegraphics[angle=0,width=0.36\textwidth]{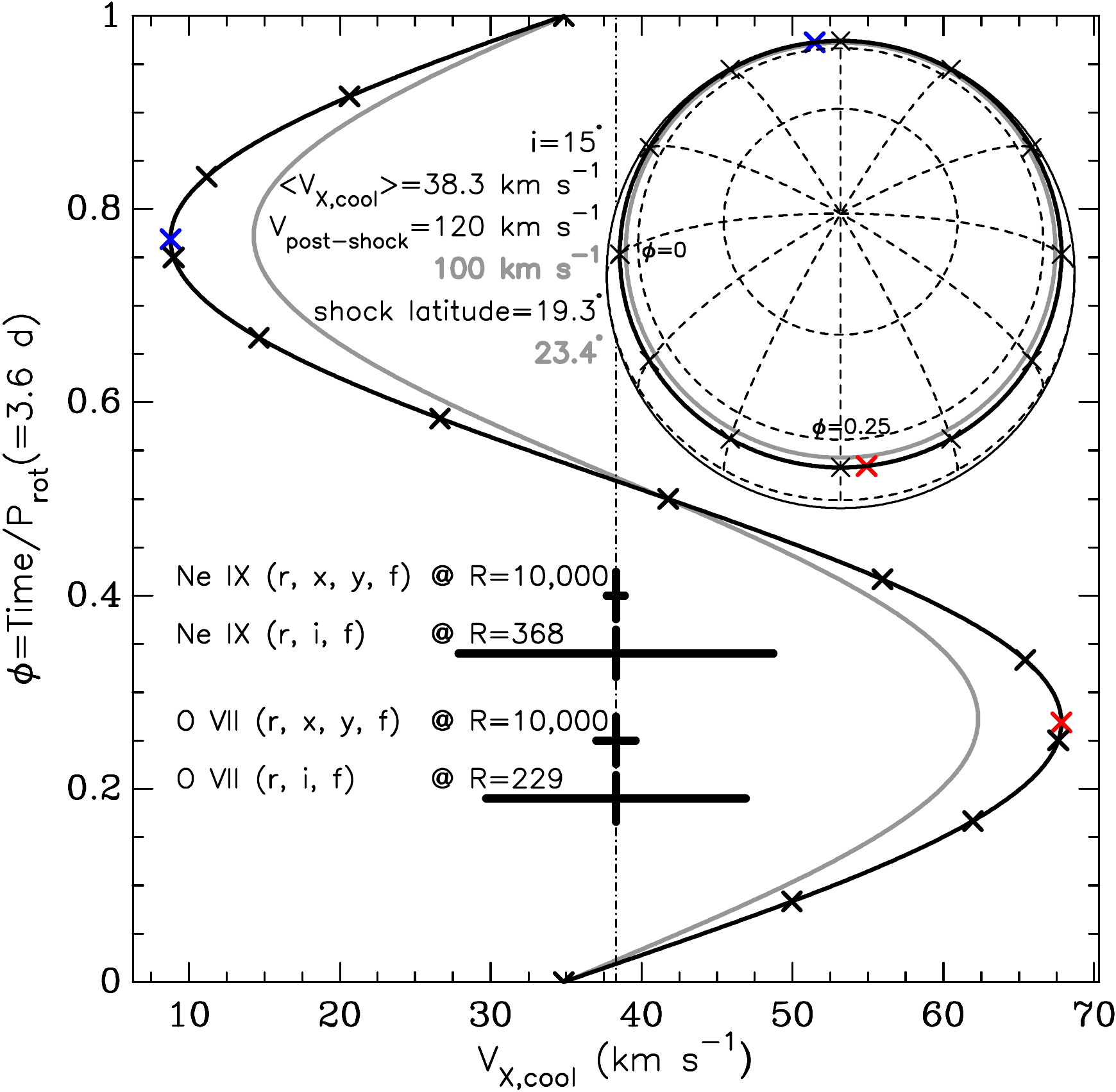}\hspace{5mm}
\includegraphics[angle=0,width=0.47\textwidth,trim={0 0 0 0.8cm},clip]{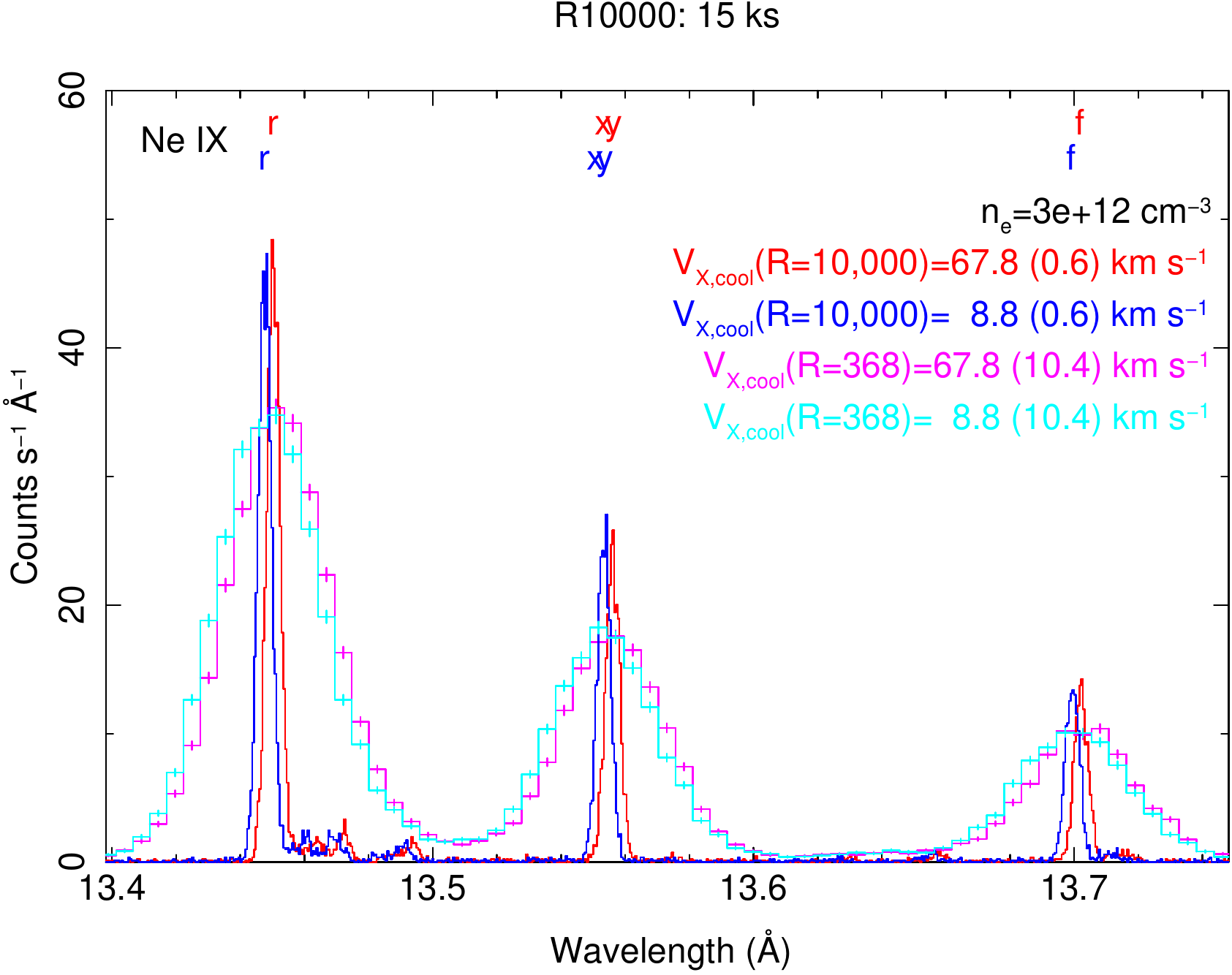}
\end{center}
\vspace{-6mm}
\caption{\footnotesize Kinematics of a point-like accretion-shock at the surface of the young accreting star TW~Hya. 
Left panel: predicted stellar rotational-phase vs. line-of-sight velocity of shocked-plasma at the accretion-stream base. 
The top-right insert displays the phase-dependent shock position on the sky-projected stellar photosphere. 
Right panel: simulated X-ray spectra of the Ne\,{\sc ix} He-like triplet with a very high-resolution high-effective area 
soft X-ray spectrometer ($\Delta\phi$=0.05, corresponding to 15\,ks), compared to Athena X-IFU \citep{BarretCappi2019}. 
The rms errors on the velocity fitted value (estimated from 500 simulated 
spectra) are provided between parentheses and as horizontal error-bars in the left panel.}
\vspace{-2mm}
\label{stars_fig1}
\end{figure} 

Low-mass stars are the most common and longest living stars. Their evolution is
substantially affected by  their exchange of mass and angular momentum with
their ambient medium during their life.
During their formation, low-mass stars accrete mass from their circumstellar
disk via magnetically channelled  streams. At the stream footpoints, the
accreting material impacts with the stellar atmosphere producing shocked  plasma
at $\sim$3\,MK. The current generation of X-ray spectrometers demonstrated that
soft X-rays are a powerful probe of the shock region
\citep[e.g.][]{GuedelNaze2009}. The shocked plasma is expected to move inward
with $v$$\sim$100\,km\,s$^{-1}$. The rotational monitoring of its line-of-sight
velocity would tightly constrain the accretion geometry. The limited spectral
resolution in soft X-rays of calorimeters will prevent these kind of studies. 
Conversely, the proposed mission concept will allow \stress{to systematically measure
the shocked-plasma velocity}  \citep[this measure is nowadays achievable only for
nearest young accreting star TW~Hya,][]{ArgiroffiDrake2017}, and to perform
\stress{Doppler-imaging of the shocked plasma on the stellar surface} for the nearest
sources (Fig.~\ref{stars_fig1}). This X-ray tomography of the extended shocked
emission will be valuable for comparison with the results of Zeemann-Doppler
imaging of stellar photospheres \citep{DonatiGregory2011}. In addition, the
\hirex mission concept will allow to resolve line widths, and hence to constrain
the turbulent  motions in the post-shock region.

When accretion ends, low-mass stars enduringly loose mass and angular momentum
via stellar winds and coronal  mass ejections (see Sect.~\ref{sect:planet}).
These phenomena are governed by the stellar magnetic activity, that is best 
studied by probing coronal phenomena in X-rays. Flares are the main
manifestation of energy transfer from the  magnetic field to the coronal plasma.
Studying the motions of plasma within flaring structures is important to
constrain flare physics, and improve coronal modeling. Plasma located near the
flaring loop footpoints (where  $T$$\sim$10\,MK and the emission measure is
maximum) is expected to move with $v$$\sim$100--500\,km\,s$^{-1}$. Such motions,
almost unexplored nowadays in stars, can be systematically studied with \hirex
by \stress{monitoring the evolution of line profiles during stellar flares.}
\athena/\xifu will likely allow to monitor line shifts (without yet resolving
line profiles) only in the extreme cases of  very hot flares occurring in
extremely active stars.

\subsection{Planets\label{sect:planet}}

\subsubsection{Exoplanet energetic photon radiation environments} 

The rate at which gas is lost from an exoplanet's atmosphere is critical for the
survivability of surface water. Atmospheric mass loss can be driven by both
thermal and non-thermal processes, which depend upon the radiation and winds of
their host stars. The dominant thermal process is hydrodynamical outflow
energised by extreme ultraviolet (EUV; 100--912~\AA) and X-radiation
(0.1--100~\AA) that heats the exoplanet's thermosphere and levitates gas against
the exoplanet's gravitational potential \citep[e.g.,][]{Owen12}.

Most of the thermospheric heating is by EUV photons but this radiation cannot be
observed directly because of interstellar H absorption. The chromospheric UV and
FUV are inadequate EUV proxies. The strength and spectral energy distribution of
a star's EUV emission instead arises from the transition region and corona. The
30--60~\AA\ range contains many lines from the same ionisation stages.
\stress{Observing these soft X-ray lines enables prediction of the EUV spectrum
and thereby constrain the atmospheric mass loss.}

Detecting the relevant lines in exoplanet hosts requires very sensitive
high-resolution ($R\geq 5\,000$) spectroscopy that is not feasible with {\textit
any existing or planned future missions}, including \chandra, \xmm or \athena.
Observations through flares and other stochastic variability are also required
to understand how EUV fluxes vary in time.

\subsubsection{Stellar winds coronal mass ejections and exoplanet atmospheric loss}

The flow of ionised stellar wind electrons and protons erode an exoplanet's
atmosphere, while coronal mass ejections can enhance the loss rate by an order
of magnitude or more \citep{Garraffo16,Dong17,Garcia-Sage17,Airapetian17}.
Recent measurements by the MAVEN satellite \citep{Brain16} confirmed that the
primary mass-loss mechanism for water on Mars is erosion by the solar wind.

The wind mass loss rates for late-type dwarfs are notoriously difficult to
measure as the solar mass-loss rate is only about $1.5\times 10^{-14}
M_{\odot}$~yr$^{-1}$. Radio observations yield only upper limits and the few
indirect estimates possible based on Ly$\alpha$ absorption in the ``wall" of
hydrogen at the stellar analogy of the heliopause \citep{Wood14} are prone to
modelling and systematic uncertainty.

Charge-exchange X-ray emission resulting from the interaction of stellar wind
ions with ISM neutral H provides a direct means of measuring wind mass loss
rates \citep{Wargelin02}. The charge exchange X-ray spectrum is dominated by
K-shell emission from H-like and He-like ions of C, O, N, and Ne. These lines
are broadened by the wind outflow velocity of 500--1000~km\,s$^{-1}$ and form
broad components underneath the narrower coronal line. A sensitive, low
background high-resolution X-ray spectrometer would be able \stress{to resolve
out this charge-exchange signal and simultaneously measure the wind velocity and
mass loss rate. This would be new stellar science not feasible at other
wavelengths.}

In addition to a relatively steady wind, stars are expected to loose mass in
coronal mass ejections (CMEs) accompanying flares. CMEs are also associated with
high-energy protons accelerated in the flare and CME shock front. CMEs are very
difficult to infer on stars and it is currently unknown how much mass and energy
are output in this way on stars other than the Sun, and what CME conditions
exoplanets experience.

\citet{Segura10} modeled the effect of a superflare ($E\approx 10^{34}$ erg) and
CME impact on a hypothetical Earth-like exoplanet located in the habitable zone
(0.16 AU) of the flare star AD Leo (dM3e). High energy protons with energies
greater than 10 MeV severely depleted nitrogen oxides, and subsequently ozone,
in the atmosphere for 2 years. \citet{Airapetian16} found CME energetic
particles can create important prebiotic molecules and alter atmospheric
greenhouse gases potentially important for the Faint Young Sun paradox.

\begin{figure}[!htb]
\includegraphics[width=0.52\textwidth]{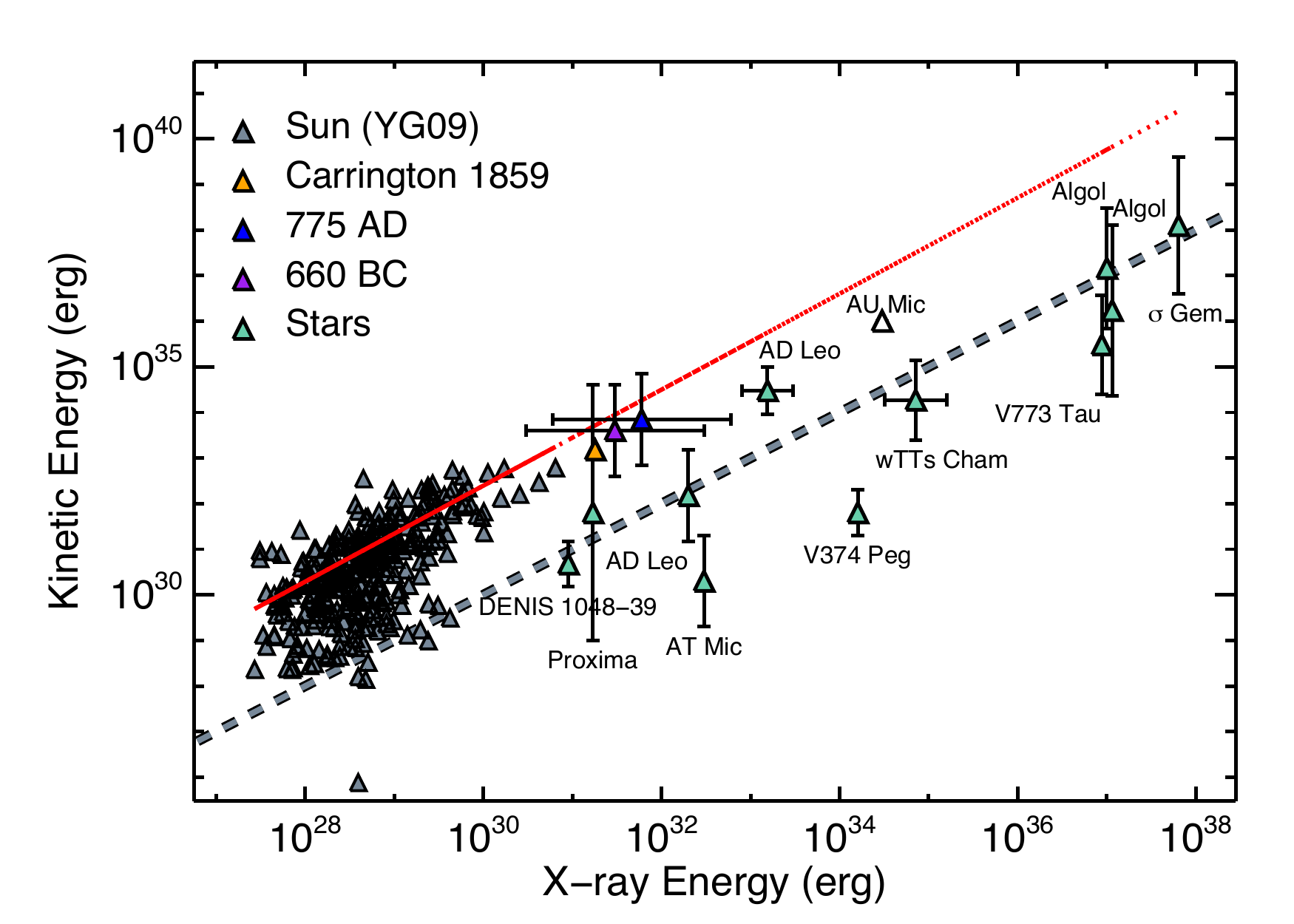}
\includegraphics[width=0.48\textwidth]{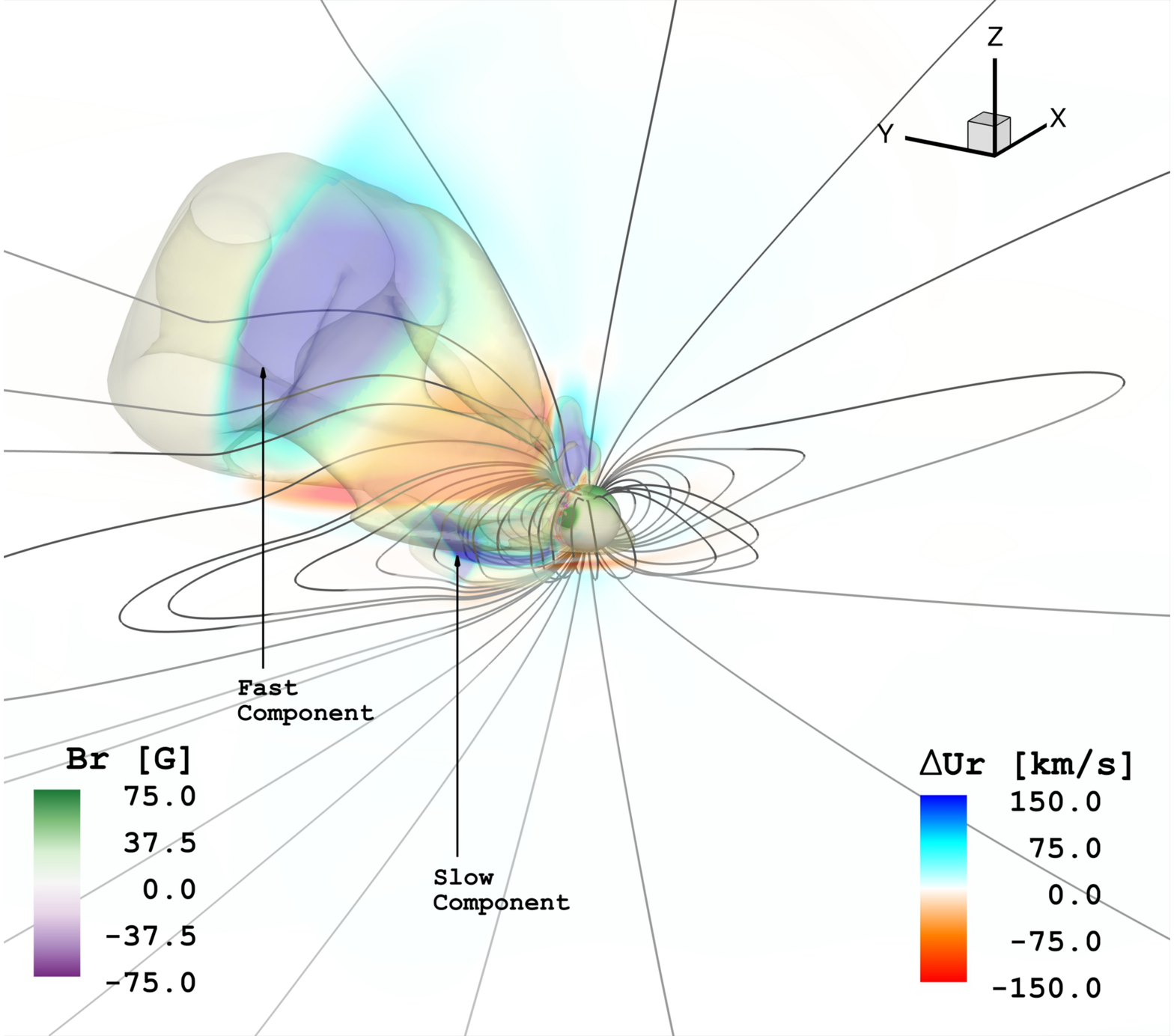} 
\caption{\footnotesize Left: The kinetic energy vs.\ associated flare X-ray
fluence for solar CMEs and CME candidate events from the literature
\citep{Moschou19}. Extrapolating the relation to large events on more active
stars is extremely uncertain, requiring definitive CME detections and
measurements for characterisation. Right: An MHD CME simulation for a moderately
active solar-like star \citep{Alvarado-Gomez18}. Plasma is compressed and
accelerated outward by the CME front, yielding {\em observable Doppler shifts},
$\Delta U_r$, of up to 100~km\,s$^{-1}$ or so. Such shifts would be detectable
with a sensitive large area grating spectrometer. 
}
\label{f:cmes}
\end{figure}

These studies demonstrate the acute need for observations of stellar CMEs.
Extrapolations of solar CME-flare relationships (Fig.~\ref{f:cmes}) are
uncertain by orders of magnitude and appear to overestimate large stellar CME
candidate kinetic energies. High-resolution spectroscopy at X-ray wavelengths
could routinely and definitively \stress{observe the tell-tale Doppler shifts of
CMEs or their coronal compression waves (Fig.~\ref{f:cmes}) and identify their
physical properties, including their thermal structure, masses and energies}, as
recently discovered by \chandra/\hetgs \citep{argiroffi2019nat}. A combination
of high throughput and high spectral resolution is critical for mapping out CME
frequency and energy vs optical and X-ray flare diagnostics for exoplanet hosts
directly, and generally as a function of spectral type and activity level.

\subsubsection{Transmission spectroscopy of exoplanet atmospheres}

X-rays are powerful diagnostics of planetary upper atmospheric gas density
structure and chemical composition. The transit of the hot Jupiter HD189733b was
detected through X-ray absorption by oxygen in \chandra observations by
\citet{Poppenhaeger13}, who found the scale height of X-ray absorbing gas was
higher than suggested by optical and UV transits. Hot Jupiters and similar giant
close-in planets are important for improving theory and models describing
atmospheric loss. 

X-ray absorption measures gas \textsl{bulk chemical composition} along the
line-of-sight --- in this case in the transiting exoplanet atmosphere backlit by
the host star's corona. Such measurements are unique to the X-ray range, but
only the very closest hot Jupiters are accessible with \chandra and \xmm, and
then only at low signal-to-noise ratio. A much more sensitive, high spectral
resolution observatory will be able to observe HD189733b-like transits out to
much greater distances, and by co-adding transits will be able to probe the
atmospheres of the nearest terrestrial planets. Absorption edge resonance
structure (not resolvable with \athena) will distinguish between atomic and
molecular or ionised gas, and provide velocity diagnostics for atmospheric
outflow. \stress{X-rays in combination with optical/IR data will provide a
powerful probe for clouds and hazes that can confuse IR spectroscopic analyses}
\citep{Sing16}.

\section{Supernova remnants} 

Nearby supernova remnants (SNRs), the outcome of SN explosions, are extended
sources which allow to study the structure and chemical composition of the
ejecta produced by the explosion. These characteristics reflect somehow the
nature of the progenitor star and pristine features of the parent SNe that may
originate from anisotropies developed at the initiation of the explosion.
Observations of SNRs, therefore, encode valuable information about the
progenitor star and the SN dynamics.

Studies of SNRs may give exceptional hints on the nature of the shocks between
the ejecta and the surrounding medium. Examples of this are the investigation of
the broadening of emission lines in ejecta-rich knots, the study of the oxygen
rich ejecta-knot of SN\,1006 performed by \citet{broersen2013} using \xmm \rgs
data, and the more comprehensive study of \citet{miceli2019} on the ion-proton
temperature ratio using \chandra gratings data of SN\,1987A.

The studies carried out so far and the already developed state-of-the-art MHD
models demonstrate the need of a technological breakthrough to increase the
spectral resolution of X-ray instrumentation in order to achieve a better
comprehension of a crucial step in the life cycle of elements in the Universe.

To clarify and quantify the required resolution, we have chosen a test case
based on real observations of the bright oxygen-rich ejecta knot of the
SN\,1006, on which an extensive analysis of grating X-ray data has been
performed, namely \xmm \rgs data \citet[]{broersen2013}, with the aim to
characterise the high resolution spectra and measure the abundances (and
therefore the masses) of different elements present in the knot. Bright X-ray
knots are ideal laboratories to study how the pristine explosion asymmetries
evolve into late-stage SNR morphologies, so the chosen test case is very
relevant for the kind of studies \hirex will be able to perform.

A 10 ks \hirex observation of the knot in SN\,1006 would allow us to gain great
insight in \stress{the chemical composition of the knot} (uncertainties $< 0.05$ in
abundances), \stress{the electron and ion temperatures of the host rich plasma,
and the turbulence and bulk flow of the knot} with uncertainties on the latter
of only 7~km\,s$^{-1}$. 

\section{Compact Objects}

Outflows are a key ingredient of accretion processes onto compact objects of all
masses. In X-ray binaries (XRBs), narrow absorption lines and P-Cygni profiles
with velocities of a few hundreds of km\,s$^{-1}$ have been observed almost
ubiquitously in high inclination sources, revealing the presence of equatorial
winds in these systems. The winds are photoionised and estimates of the expelled
mass indicate that it could be large enough to trigger accretion state changes
or even be the reason for an outburst to cease
\citep{shields1986,ponti2012,munoz2016}. A range of ionisations is observed
whenever sensitive observations are available, indicating some stratification in
the wind \citep[e.g.,][]{ueda2004,miller2006,kallman2009}. In particular, a low
ionisation component of the wind is present in the majority of XRBs that are not
absorbed below $\sim$2 keV in the interstellar medium \citep[see Table 1
in][]{diaz2016}. 

Resolving the line profiles provides us with a powerful diagnostic for
characterising the winds in XRBs \citep{ueda2004,kaastra2014}. Line profiles are
extensively used at UV, optical, IR or mm wavelengths where resolving powers
well above 3000 are available. For example, \citet{calvet1993} used them to
identify an accelerating wind in pre-main sequence objects as originating from
the accretion disc. In addition, the region between 0.2--2 keV contains
diagnostic lines for plasma density like \ion{Fe}{xvii} and \ion{Fe}{xxii}
\citep{mauche2004}, a crucial parameter to determine the distance between the
plasma and the ionising source, which is key to discern among different wind
launching mechanisms. Moreover, the wealth of lines in this energy band is also
fundamental to diagnose if a plasma is collisionally ionised or photoionised,
and thus to differentiate between absorption in the interstellar medium or local
to the source. 

\begin{figure}[!htb]
\vspace{-4cm}
\begin{center}
\includegraphics[angle=0.0,width=0.7\textwidth]{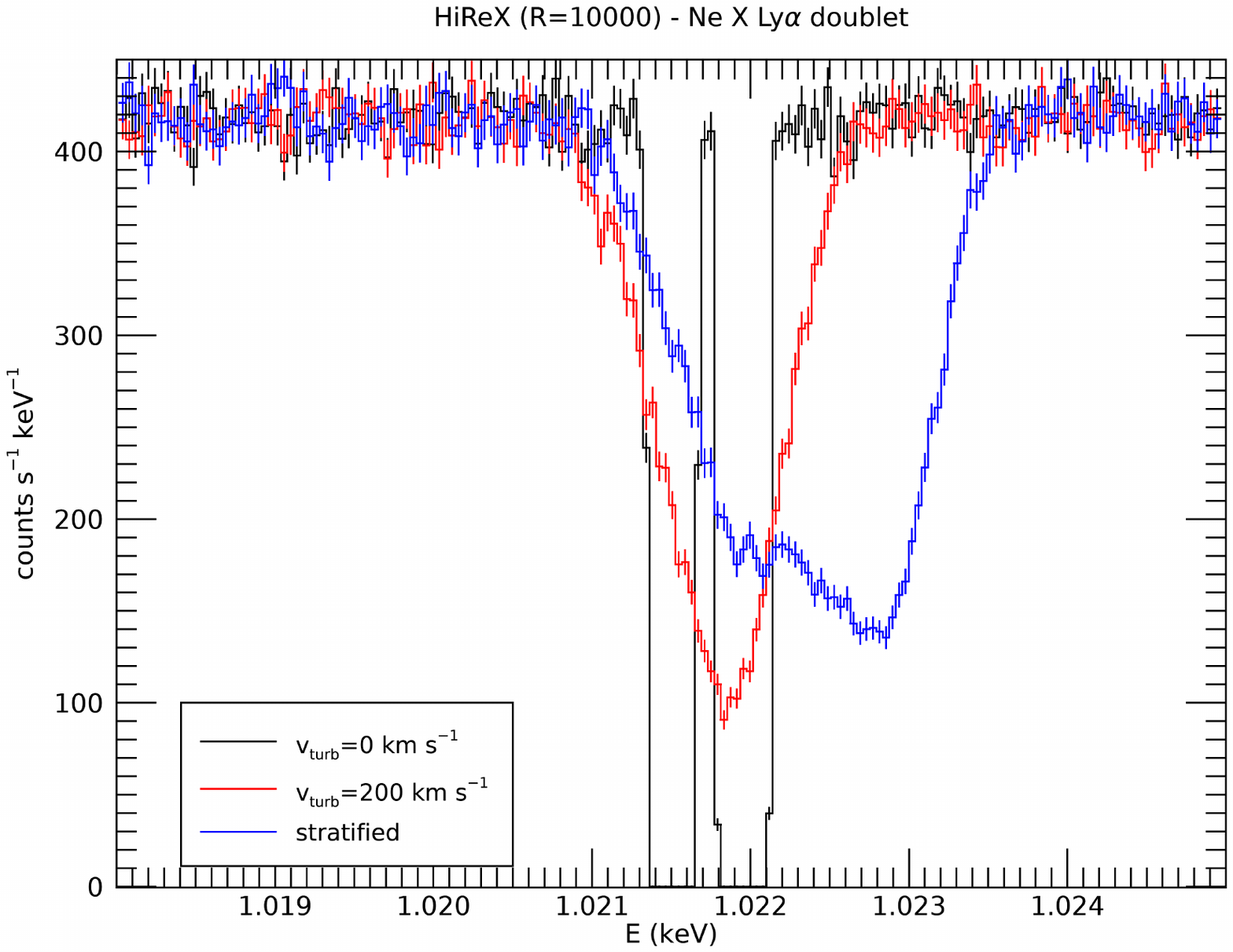}
\vspace{-4.5cm}
\caption{\footnotesize \ion{Ne}{x} line profiles expected for a a thermal wind
with no significant turbulence (black), a thermal wind with a turbulence of 200
km\,s$^{-1}$ (red) and a stratified wind with a velocity profile between 0 and
300 km\,s$^{-1}$ (blue) and no turbulence. At a resolution lower than $\sim$0.2
eV, the line profiles in black and red will be indistinguishable due to the
blend of the line doublet. However, with the resolving power of \hirex, we will
be able to resolve the velocity profile of the wind (blue) allowing us to make
detail comparisons with the expectations for thermal, radiative or magnetic
winds and to distinguish between a profile with or without turbulence. All
diagnostics shown will critically depend on the capability to recognise a
saturated line (see text), which otherwise would introduce an unknown systematic
uncertainty in the resolution and interpretation of the line profiles.}
\end{center}
\label{fig:wind}
\end{figure}

\hirex will allow a study of the low ionisation wind components at unprecedented
resolution in X-rays. For the first time we will be able \stress{to resolve the line
forest between 14--18 \AA, determine turbulent and thermal broadening of the
lines and measure radial velocities even in the smallest systems}. Only resolving
powers of $\sim$5000 or better, as provided by \hirex below 2~keV, will allow us
to disentangle line saturation from turbulent and thermal broadening or a blend
of the two components of a H-like ion doublet (see Fig.~\ref{fig:wind}). This
and the large amount of lines present in the 0.2--2~keV region will allow us to
choose non-saturated lines to map the accretion disc atmospheres and winds with
exquisite detail. The line spectra from atmospheres (dominated by lines from
He-like and H-like elements from C to Fe, radiative recombination lines and Fe L
lines) are very sensitive to temperature, ionisation and emission measure of
each atmospheric layer, probing the heating mechanisms in the disc
\citep{jimenez2002}. For winds, thermal pressure
\citep{begelman1983a,begelman1983b}, aided by radiation pressure for systems
with luminosities above 50\% Eddington, is currently favoured as the main
launching mechanism, but the topic of whether there is room for a magnetic wind
or not is highly debated \citep[e.g.][]{done2018,fukumura2018,waters2018}. The
high resolution line profiles obtained with \hirex will allow us to disentangle
between these mechanisms since the wind velocity/acceleration profile is
expected to be different for thermal, radiative and magnetic pressure winds
\citep{ueda2004,fukumura2010}. In this sense, \stress{\hirex will be complementary to
\athena by expanding the studies of the highest ionisation component of winds to
the lower ionisation ones, but adding fundamental diagnostic parameters such as
density.}

While \athena will provide a resolving power close to 3000 at 7~keV, sufficient
to resolve line profiles for the most ionised component of the plasma revealed
by the presence of \ion{Fe}{xxv} and \ion{Fe}{xxvi} lines, at energies below
2~keV, only \hirex provides the resolving power needed to make the same studies
for the lower ionisation components of the wind. Note that the \ion{Ne}{x}
doublet is only separated by $\sim$0.5~eV, compared to $\sim$20~eV for the
\ion{Fe}{xxvi} doublet \citep{verner1996}. The recently reported presence of
optical winds during the accretion state where jets are observed
\citep[e.g.,][]{rahoui2014,munoz2016} signals the presence of relatively cool
material that could be associated to the soft X-ray wind component. Since the
high ionisation wind component is absent in such an accretion state and only
appears at states dominated by strong thermal emission, the mechanisms behind
the X-ray low and high ionisation components could be different. \hirex will
allow us \stress{to fully characterise the low ionisation component of winds and test if
the launching mechanism is the same as for the high ionisation winds}. Finally,
the line profiles will also allow us to determine the mass loss rate in the wind
by constraining the solid angle of the wind through the ratio of emission to
absorption.

\section{A possible mission concept\label{sect:concept}}

Resolving the thermal Doppler width (FWHM) of a transition in an ionic species
of atomic mass $M$ in a plasma of temperature $T$ requires a resolving power
${R} = 9700 (M/56)^{1/2} (T/10^6\ {\rm K})^{-1/2}$. We choose ${R} =
10\,000$ as our reference value. At this value, we resolve the lines of Fe ions,
the heaviest abundant element, in essentially all X-ray plasmas. Given
sufficient photon flux, such resolving power will also enable the measurement of
bulk velocities down to a few km\,s$^{-1}$.

To maintain an approximately constant resolving power across an extended X-ray
energy band, diffractive spectrometers offer the best combination of flexibility
and ease of practical implementation. The resolving power of a diffraction
grating spectrometer scales approximately inversely proportional to photon
energy, that of a microcalorimeter-based spectrometer approximately proportional
to photon energy. But a diffraction grating spectrometer can be operated in a
series of different spectral orders, essentially as an 'echelle' spectrograph,
which allows for a more uniform resolving power coverage of the chosen band. We
explore a simple design based on laboratory-proven technology.

Two types of diffraction grating are currently being developed for
high-resolution astrophysical X-ray spectroscopy: the Critical Angle
Transmission grating \citep{heilmann2019}, and the 'off-plane' radial groove
gratings \citep{mcentaffer2019}. Very roughly speaking, the first type combines low
tolerance alignment implementation with a relatively meticulous manufacturing
process, while the second type is faster to manufacture but requires precise
optical alignment. Both types offer a natural combination of high dispersion and
high diffraction efficiency. We detail a possible design based on CAT gratings.

We choose 100--2000 eV as the baseline energy range. The energy range 500--1000
eV (12--23 \AA) is especially rich in astrophysically important transitions (the
O K-shell and Fe L shell species), and we blaze the spectrometer at 15 \AA. For
a grating period $d = 2000$ \AA, the angle of incidence on Si grating bars
should not exceed 2 degrees, and the spectrometer operates in spectral orders $m
\approx 9$ around the blaze wavelength \citep{heilmann2019}. Order separation
can be ensured by reading out the spectroscopic image with a semiconductor
imaging detector with an energy resolving power of order 10 or better, such as
delivered by a Fano-limited Si device. Assuming a focusing optic with angular
resolution $\leq 1$ arcsec, and ignoring spherical aberration and the effect of
grating imperfections, results in a spectroscopic resolving power at blaze of
14\,400. Spectral orders 1--12 provide a resolving power over 10\,000 at all
wavelengths longer than about 8 \AA.

If we assume that systematic effects (uncalibrated pixel-to-pixel quantum
efficiency variations) in the focal plane detector are the ultimate limit to the
minimum detectable line equivalent width, then, for an assumed 3\% systematic
error, the minimum detectable equivalent width is 3\% of the instrument
resolution, or $EW_{\rm limit} = 3 \times 10^{-2} (\lambda/10\,000) = 3 \times
10^{-6} \lambda$ \AA = $0.06 (\lambda/20$\AA) m\AA.

\begin{table*}[!htbp]
\caption{\footnotesize Figure of Merit (FoM) for measuring equivalent widths of weak lines
at 0.5~keV energy for various high-resolution X-ray spectrometers. 
The expected detection significance scales with FoM $\sim\sqrt{A_{\rm eff}R}$
with $A_{\rm eff}$ the effective area and $R$ the resolution of the instrument. }
\label{tab:fom}
\smallskip
\centerline{
\begin{tabular}{lcccc}
\hline\hline\noalign{\smallskip}
Mission & Instrument & $A_{\rm eff}$ (cm$^2$) & $R$ & FoM \\
\hline\noalign{\smallskip}
\hirex & & 1\,500 & 10\,000 & $\equiv$ 1 \\
\chandra & \letgs & 12 & 500 & 0.02 \\
\xmm & \rgs & 90 & 400 & 0.05 \\
\xrism & \resolve & 125 & 100 & 0.03 \\
\athena & \xifu & 5\,900 & 200 & 0.28 \\
\hline\noalign{\smallskip}
\end{tabular}
}
\end{table*}

Table~\ref{tab:fom} compares the figure of merit (expected significance) of weak
line detections for various instruments. In this respect \hirex\ is 20--50 times
better than the present grating spectrometers. While \athena/\xifu can
compensate the lower spectral resolution by more effective area in terms of
statistical significance of line detections, the ultimate limit for weak lines
is determined by the systematic uncertainties outlined above. Assuming a typical
3\% systematic uncertainty for all these instruments, the ultimate performance
is then determined solely by the resolving power $R$. \stress{In this respect, for
detection of weak lines, \hirex is
better than all other instruments by a factor of 20--50.}

\section{Acknowledgements}

Team members thank the following people for supporting this effort and
contributing to the discussion and development of the paper: 
M. Audard (Un. of Geneva), 
M. Barbera (INAF), 
D. Barret (CNRS - IRAP, France), 
G. Betancourt-Martinez (IRAP), 
V. Biffi (INAF/SAO),
E. Branchini (Un. Roma Tre), 
E. Costantini (SRON), 
J.R. Crespo (MPI-Heidelberg),
K. Dolag (MPA),
A. Finoguenov (Un. of Helsinki), 
V. Grinberg (Un. of Tuebingen),  
J.W. den Herder (SRON), 
I. Khabibullin (MPA \& IKI),
A. Maggio (INAF), 
Y. Naz\'e (University of Li\`ege), 
S. Paltani (Un. of Geneva), 
L. Piro (INAF), 
G. Ponti (MPE), 
E. Rasia (INAF), 
G. Rauw (University of Li\`ege), 
F. Reale (INAF), 
A. Simionescu (SRON),
J. Wilms (R. Obs. Bamberg).


\newpage 

\bibliography{hirex}

\newpage

\begin{center}
{\Large {\bf{Members of the Core Proposing Team}}} 
\end{center}
\vspace{2cm}

\noindent\newline F. Nicastro (INAF - Osservatorio Astronomico di Roma, Italy),
\noindent\newline J. Kaastra (SRON \& Leiden University, The Netherlands),
\noindent\newline C. Argiroffi (University of Palermo, Italy), 
\noindent\newline E. Behar (Technion, Israel), 
\noindent\newline S. Bianchi (Universit\'a degli Studi Roma Tre, Italy), 
\noindent\newline F. Bocchino (INAF - OAPA, Italy), 
\noindent\newline S. Borgani (University of Trieste, Italy), 
\noindent\newline G, Branduardi-Raymont (MSSL, UK), 
\noindent\newline J. Bregman (University of Michigan, USA), 
\noindent\newline E. Churazov (MPA, Germany \& IKI, Russia), 
\noindent\newline M. Diaz-Trigo (ESO, Germany), 
\noindent\newline C. Done (University of Durham, UK), 
\noindent\newline J. Drake (SAO - CfA, USA), 
\noindent\newline T. Fang (Xiamen University, China), 
\noindent\newline N. Grosso (Aix Marseille University, CNRS, CNES, LAM, France), 
\noindent\newline  A. Luminari (University of Rome "Tor-Vergata", Italy  \& INAF - OAR), 
\noindent\newline M. Mehdipour (SRON, The Netherlands), 
\noindent\newline F. Paerels (Columbia University, USA), 
\noindent\newline E. Piconcelli (INAF - OAR, Italy), 
\noindent\newline C. Pinto (ESA, The Netherlands), 
\noindent\newline D. Porquet (Aix Marseille University, CNRS, CNES, LAM, France), 
\noindent\newline J. Reeves (University of Maryland, USA), 
\noindent\newline J. Schaye (Leiden Observatory, The Netherlands), 
\noindent\newline S. Sciortino (INAF - OAPA, Italy), 
\noindent\newline R. Smith (SAO - CfA, USA),
\noindent\newline D. Spiga (Stanford University - SLAC-LCLS, USA),
\noindent\newline R. Tomaru (University of Tokyo, Japan), 
\noindent\newline F. Tombesi (University of Rome "Tor-Vergata", Italy),  
\noindent\newline N. Wijers (Leiden Observatory, The Netherlands),
\noindent\newline L. Zappacosta (INAF - OAR, Italy)


\end{document}